\documentclass[sigconf,10pt,screen]{acmart}

\newif\ifnotnewacm
\notnewacmfalse

\usepackage{booktabs}
\usepackage{tabularx}
\usepackage{textcase}
\usepackage{listings}

\usepackage[T1]{fontenc}
\usepackage[scaled=0.85]{FiraMono}

\AtBeginDocument{
    
}

\renewcommand{\texttt}[1]{{\ttfamily\footnotesize #1}}

\definecolor{codeblue}{HTML}{0000FF}      
\definecolor{codepurple}{HTML}{AF00DB}    
\definecolor{codeteal}{HTML}{267F99}      
\definecolor{codebrown}{HTML}{A31515}     
\definecolor{codedarkblue}{HTML}{001080}  
\definecolor{codeyellow}{HTML}{795E26}    
\definecolor{codegreen}{HTML}{008000}     
\definecolor{codegray}{HTML}{808080}      

\usepackage{caption}
\newcommand{\ncaption}[1]{%
  \caption{\textnormal{#1}}%
}

\makeatletter
\def\@ACM@checkaffil{
    \if@ACM@instpresent\else
    \ClassWarningNoLine{\@classname}{No institution present for an affiliation}%
    \fi
    \if@ACM@citypresent\else
    \ClassWarningNoLine{\@classname}{No city present for an affiliation}%
    \fi
    \if@ACM@countrypresent\else
        \ClassWarningNoLine{\@classname}{No country present for an affiliation}%
    \fi
}
\makeatother

\newif\ifarxiv
\arxivtrue 

\newif\ifheadnice

\ifarxiv
\headnicefalse
\else
\headnicetrue
\fi

\def\sys{\textsc{TensorCast}\xspace}

\renewcommand\footnotetextcopyrightpermission[1]{} 

\usepackage[T1]{fontenc}

\title{\sys: The Missing Tensor Management\\ Layer in Large Language Model Infrastructure}

\def\pku{\superscript{$\mathbb{P}$}}
\def\stepfun{\superscript{$\mathbb{S}$}}
\def\bupt{\superscript{$\mathbb{B}$}}

\begin{CJK}{UTF8}{gbsn}

\newtheorem{algorithm}{算法}

\title{***}
\author{***footnote{电子邮件:**}\\[2ex]
*** \\[2ex]
}
\date{20XX年X月}

\maketitle

\tableofcontents
\newpage
在此输入正文，中英文均可。

\songti 
\fangsong 
\biaosong
\heiti
\kaishu

\end{CJK}

\usepackage[font=normal, skip=2pt]{caption} 

\author{
\large {Yuhan Zhou\pku\footnotemark[1], Yuchu Luo\footnotemark[1], Hao Nie\stepfun, Wangrunze Lv\bupt\footnotemark[1], Yu Zhou\stepfun, Yibo Zhu\stepfun, Daxin Jiang\stepfun, Chenren Xu\pku}
\\
\large
\begin{tabular}{ccc}
{\pku}Peking University & {\stepfun}Stepfun & {\bupt}Beijing University of Posts and Telecommunications
\end{tabular}
}

\begin{document}
    \fancyhead{}
    \begin{abstract}

Modern LLM infrastructure increasingly manages tensors not only as
computation data, but also as persistent states shared across
distributed components. Existing systems optimize individual tensor
management tasks, such as model weight loading, KV cache management,
and checkpoint synchronization, by deeply integrating task-specific
mechanisms with execution engines, networks, or storage backends.
However, this specialization creates isolated silos that hinder the
reuse and composition of tensor management strategies across evolving
LLM workloads.
In this paper, we identify tensor lifecycle management as a missing
abstraction layer in LLM infrastructure and propose Tensor-as-a-Service
(TaaS), which decouples tensor state management from computation logic.
We design and build \sys, a distributed tensor management layer that
provides first-class tensor abstractions, programmable lifecycle
primitives, and a runtime that separates tensor management policies from
execution mechanisms. This enables developers to write tensor
management programs using \sys APIs while transparently leveraging
distributed execution and data movement.
We integrate \sys with vLLM and SGLang and evaluate it across diverse
tensor lifecycle workloads, including model weight materialization,
weight synchronization, KV cache management, and programmable request
routing. Our results show that \sys achieves competitive performance
with specialized tensor management systems while enabling new
cross-component optimization policies. A programmable policy implemented
with \sys improves median TTFT by up to 93.2\% under highly concurrent multi-turn agent workloads.

\end{abstract}

    \maketitle
    \begingroup
        \renewcommand{\thefootnote}{\fnsymbol{footnote}}
        \footnotetext[1]{This work was done while Yuhan Zhou, Yuchu Luo, and Wangrunze Lv were affiliated with Stepfun as employees or interns.}
    \endgroup

\section{Introduction}

Modern LLM infrastructure is rapidly scaling in both model size and
deployment complexity, driven by models with billions to trillions of
parameters (\eg 2.8T parameters in Kimi K3 \cite{kimi-k3}) and stringent
Service Level Objectives (SLOs) for millions of users. This evolution
has pushed LLM systems beyond optimizing tensor computation alone:
tensors increasingly become persistent states that must be managed across distributed components. For example, model weights are
distributed across elastic serving instances, KV caches are reused
across nodes, and updated model weights are synchronized
between training and inference pipelines.

\begin{figure}
	\centering
	\includegraphics[width=\linewidth]{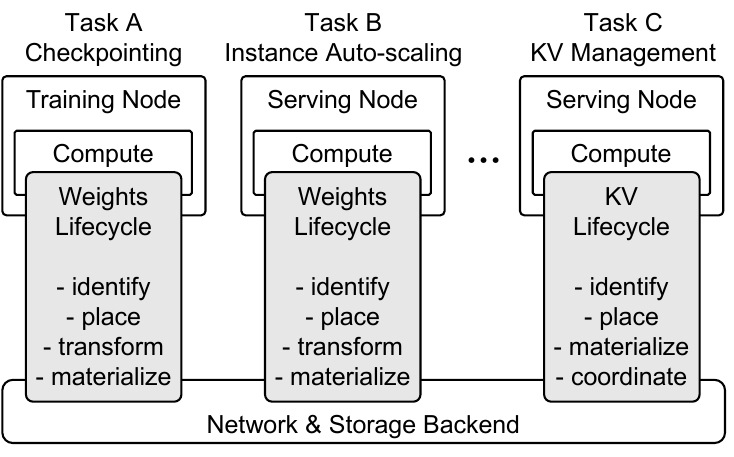}
	\caption{Existing LLM infrastructure embeds tensor lifecycle management into task-specific stacks, causing repeated mechanisms and isolated optimization silos.}
	\label{fig:vertical-silo}
\end{figure}

\begin{figure}
	\centering
	\includegraphics[width=\linewidth]{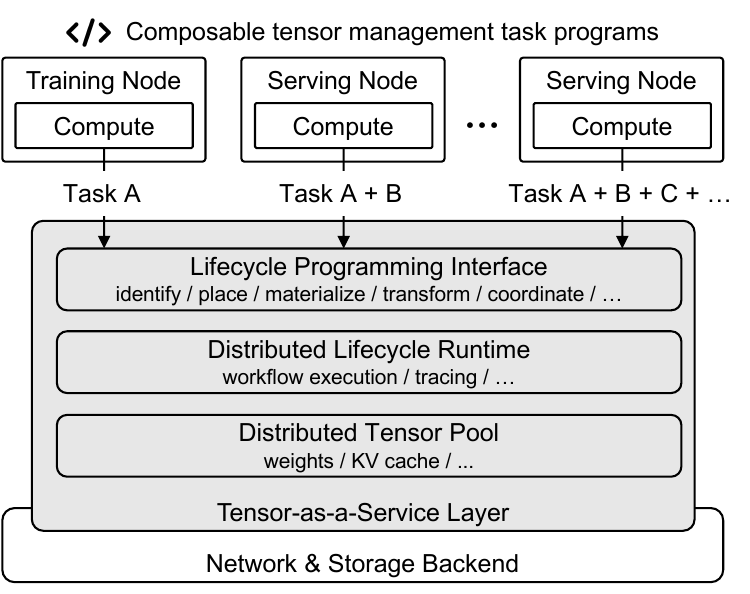}
	\caption{Decoupling tensor lifecycle management from
computation. A unified Tensor-as-a-Service (TaaS) layer allows developers to compose tensor management task programs across workloads.}
	\label{fig:tensor-management-layer}
\end{figure}

A large body of research has optimized individual tensor management
tasks in LLM infrastructure. Representative examples include model
weight loading for service auto-scaling
\cite{fu2024-serverlessllm, hu2025-deepflow, yu2023-faaswap,
yu2025-lambdascale, zhang2025-blitzscale, lou2025-hydraserve,
sglang-rfork}, KV cache placement and transfer for prefill-decode
disaggregation \cite{qin2025-mooncake, anonymous2025-dualmap} and prefix
cache sharing \cite{liu2025-lmcache, sglang-hicache,
yu2025-pensieve, yao2025-cacheblend, gao2024-cachedattention}, as well
as model checkpoint storage and resharding for large-scale training
\cite{wan2025-bytecheckpoint, moonshot-ckpt-engine}. These systems
achieve significant performance improvements by tightly integrating
tensor management mechanisms with execution engines
\cite{vllm, zheng2024sglang, tensorrt-llm}, network infrastructures
(\eg scale-up versus scale-out networks
\cite{nvidia-scalable-network}), or storage backends.

However, this task-specific optimization paradigm creates a fundamental
abstraction gap in LLM infrastructure. Although different workloads
manipulate tensors through similar lifecycle operations such as
identifying, placing, moving, transforming, and materializing tensor
states, existing systems expose these operations only through
workload-specific interfaces tightly coupled with execution engines,
network layers, or storage backends, creating isolated vertical silos
(\figref{fig:vertical-silo}).  Such tight coupling makes the underlying tensor management mechanisms difficult to extract, reuse, or compose across workloads.
This lack of a unified tensor lifecycle abstraction leads to two
fundamental limitations. First, developers repeatedly reimplement
similar tensor lifecycle mechanisms for each workload, increasing system
complexity and slowing the evolution of LLM infrastructure. For
example, adapting a serving scheduler from a load-aware policy to a
cache-affinity-aware policy may require coordinated modifications
across request routers, KV cache systems, and execution engines.
Second, the lack of composability prevents LLM systems from exploiting
optimization opportunities that span multiple components. A distributed
LLM service may need to simultaneously balance instance load, preserve
KV cache locality, and migrate request states across instances. However,
existing siloed solutions make such cross-component optimization
difficult to express.

In this paper, we argue that \textit{tensor management should become an
independent abstraction layer rather than remain embedded inside
individual execution systems}. As tensor management requirements
continue to evolve and often require composing multiple lifecycle
operations, such a layer must provide programmable control over tensor
states throughout their lifecycle. We propose Tensor-as-a-Service
(TaaS), a new abstraction layer that decouples tensor lifecycle
management from the computation that produces or consumes tensors
(\figref{fig:tensor-management-layer}).
Guided by this vision, we design and build \sys, a distributed tensor
management layer for LLM infrastructure. \sys realizes TaaS through
three key design principles:
\textit{i) Tensor-native abstraction.}
\sys represents tensors as first-class system objects with explicit
identity, ownership, and lifecycle semantics, enabling developers to
manage tensors independently from their physical placement and
representation;
\textit{ii) Programmable lifecycle management.}
\sys exposes composable primitives for tensor movement, transformation,
and materialization, allowing developers to define workload-specific
policies without reimplementing low-level mechanisms;
\textit{iii) Policy-mechanism separation.}
\sys separates tensor management policies from execution mechanisms:
developers can write ordinary programs using \sys APIs to compose and
control tensor lifecycles, while the \sys runtime transparently executes
these operations across distributed cluster resources without requiring
modifications to the underlying execution engines.
Unlike traditional distributed object stores that treat data as opaque
blobs \cite{redis, yu2023-vineyard, ray-plasma}, or computation frameworks that schedule user-defined tasks \cite{moritz2018-ray, zaharia2012-spark},
\sys provides a data-centric programming model where developers
control tensor lifecycle policies while the system manages the execution mechanisms, including placement, movement, transformation, and materialization.

We integrate \sys with vLLM \cite{vllm} and SGLang \cite{zheng2024sglang} LLM inference frameworks, and evaluate it
across diverse tensor lifecycle scenarios. Our results demonstrate that \sys achieves competitive performance with specialized tensor management systems in model weight materialization (\secref{sec:eval-loader}), weight synchronization (\secref{sec:eval-updater}), and KV cache management (\secref{sec:eval-kv}), while enabling new cross-component optimization policies through programmable lifecycle composition. In particular, a programmable request router implemented with
\sys reduces median TTFT by up to 93.2\% under highly concurrent multi-turn agent workloads by composing tensor lifecycle operations to jointly balance instance load and KV cache locality without modifying existing serving components (\secref{sec:eval-router}). \sys is open-sourced at \url{https://github.com/tensorcast-ai/tensorcast}.

\nosection{Contributions.}
\begin{Itemize}
    \item We identify tensor lifecycle management as a missing
    abstraction layer in modern LLM infrastructure and characterize the
    common lifecycle primitives shared across diverse tensor management
    workloads.

    \item We design and build \sys, a distributed tensor management
    layer that provides first-class tensor abstractions and programmable lifecycle primitives to separate management policies from execution mechanisms.

    \item We implement \sys and evaluate it across model weight
    materialization and synchronization, KV cache management, and
    programmable request routing. The results demonstrate both the
    efficiency and programmability of the proposed tensor lifecycle
    abstraction.
\end{Itemize}

\textbf{This work does not raise any ethical issues.}

\section{Background and Motivation}\label{sec:bkgd-moti}

\subsection{Tensor Management Tasks in LLM Infrastructure}\label{sec:bkgd}
LLM training and serving workloads increasingly treat tensors not only as intermediate computation results, but also as persistent states that must be created, transferred, stored, and transformed across system components. Representative examples include model weights shared across serving instances, KV caches reused across requests, and checkpoint tensors synchronized across training and inference workers. Consequently, efficient management of these tensor states becomes a fundamental challenge in LLM infrastructure. We next discuss three representative scenarios.

\nosection{Service Auto-Scaling}
To accommodate fluctuating request loads, LLM service providers dynamically scale serving instances, enabling a serverless Model-as-a-Service (MaaS) paradigm \cite{mcgrath2017-serverless, yang2022-infless}. Unlike traditional model deployment where weights are loaded once during initialization, elastic LLM serving requires repeatedly distributing large model weights (often exceeding 100 GB) to newly launched instances. Therefore, model weights become shared tensor states whose placement and materialization directly affect scaling latency. Existing solutions optimize this
process by streamlining loading from storage backends \cite{fu2024-serverlessllm, hu2025-deepflow, yu2023-faaswap} or retrieving weights from active serving instances \cite{yu2025-lambdascale, zhang2025-blitzscale, lou2025-hydraserve, sglang-rfork}.

\nosection{KV Cache Management.}
Modern Transformer-based LLM serving systems \cite{vaswani2017-attention, brown2020-lm} maintain KV caches to avoid recomputing attention states during autoregressive generation. These KV tensors, generated throughout the prefill and decode phases, preserve the context required for future token generation. As context lengths grow and multi-turn interactions become prevalent, KV caches become
large, fragmented, and reusable tensor states rather than transient execution buffers. Consequently, substantial research focuses on KV cache placement and transfer \cite{qin2025-mooncake, anonymous2025-dualmap, zheng2024sglang, gao2024-cachedattention, hu2025-flexkv} for prefill-decode
disaggregation \cite{zhong2024-distserve, patel2024-splitwise, hu2024-tetriInfer, strati2024-dejavu}, as well as KV reuse strategies for prefix caching \cite{gim2024-promptcache, liu2024-cachegen, yao2025-cacheblend, liu2025-lmcache, jin2025-ragcache, gao2024-attentionstore}.

\nosection{Dynamic Checkpointing and Resharding.}
During LLM training and post-training, model tensors frequently move between different system components and parallelism configurations. Checkpointing periodically materializes model states for fault tolerance and evaluation, while RL post-training requires synchronizing newly updated weights from training actors to rollout workers. Furthermore, heterogeneous GPU resources and changing parallelism strategies require resharding checkpoint tensors into different layouts \cite{wan2025-bytecheckpoint}. Consequently, several specialized systems have emerged to optimize model checkpointing, synchronization, and resharding processes \cite{wan2025-bytecheckpoint, moonshot-ckpt-engine, megatron-dcp, pytorch-dcp, lian2024-ucp}.

\begin{table*}[htbp]
    \centering
    \begin{tabularx}{\textwidth}{l l >{\hsize=1.1\hsize}X >{\hsize=0.9\hsize}X}
        \toprule
        \textbf{Workload/Task} & \textbf{Tensor State} & \textbf{Lifecycle Challenge} & \textbf{Common Lifecycle Primitives} \\
        \midrule
        MaaS serving
         & Model weights
         & Distribute immutable weights to new instances.
         & identify, place, transform, materialize \\
        KV cache management
         & KV cache blocks
         & Share reusable context across requests/instances.
         & identify, place, materialize \\
        RL post-training
         & Model weights
         & Synchronize updated weights from trainers to rollout workers.
         & identify, transform, materialize \\
        Agentic reasoning
         & Reasoning states
         & Branch, offload, and restore states across instances.
         & identify, materialize, coordinate\\
        \bottomrule
    \end{tabularx}
    \caption{Common tensor lifecycle abstractions across representative LLM infrastructure workloads.}
    \label{tab:tensor-lifecycle}
\end{table*}

\subsection{Observed Tensor Lifecycle Primitives}\label{sec:lifecycle-primitives}

Although these workloads differ in application semantics, they expose a common feature: tensor states are repeatedly manipulated through a small set of common lifecycle primitives that are shared across different application logic. Tensors require independent management of identity, placement, transformation, and coordination beyond the computation that produces or consumes them, as summarized in \tabref{tab:tensor-lifecycle}.

\begin{Itemize}
    \item \textbf{Identify and Own.} Tensors first need stable identity, metadata, and ownership semantics before they can be shared across engines, processes, or time;
    \item \textbf{Place and Move.} Tensor states must be placed, replicated, migrated, or prefetched across nodes and instances according to workload policies;
    \item \textbf{Materialize.} A logical tensor state must eventually be materialized into a compute-ready representation inside an execution engine;
    \item \textbf{Transform.} Tensor states often require slicing, viewing, layout conversion, or resharding when crossing parallelism configurations or engine boundaries;
    \item \textbf{Compose and Coordinate.} Real workloads combine multiple lifecycle steps, such as publishing, moving, transforming, and materializing tensors, and therefore require a policy-level mechanism to coordinate these operations across distribute components.
\end{Itemize}
These observations reveal that tensor management is no longer tied to individual workloads or execution engines. Instead, LLM infrastructure requires a common abstraction for managing tensor states across their lifecycles, from ownership to movement, transformation, and coordination.

\subsection{Limitations of Specialized Solutions}

Recently, numerous dedicated solutions have emerged to optimize individual tensor management tasks within LLM infrastructure, bringing significant performance improvements, as discussed in \secref{sec:bkgd}.
However, despite sharing these common lifecycle primitives, existing systems expose them only through
task-specific interfaces. Consequently, the same underlying tensor operations are repeatedly reimplemented and deeply coupled with different execution engines, network layers, and
storage backends. This fragmentation leads to the following limitations.

\nosection{Lifecycle primitives are coupled with task-specific mechanisms.}
Specialized solutions often integrate tensor lifecycle operations directly with workload-specific optimization logic. For example, Mooncake~\cite{qin2025-mooncake} adopts a min-TTFT strategy to schedule requests on prefill instances, which requires estimated prefill durations from the execution engine and predicted KV transfer times from the network. While such tight integration enables effective
optimization for KV-centric serving, the underlying tensor placement and movement mechanisms cannot be easily reused by other workloads.
Consequently, tensor management tasks such as model weight distribution and KV cache migration are implemented as independent ``vertical silos'' that are deeply integrated with execution engines, network layers, and storage backends (\figref{fig:vertical-silo}). This task-specific design
creates substantial engineering overhead as new tensor lifecycle requirements emerge. Furthermore, because optimization policies are frequently hard-coded within individual systems, adapting them requires modifying low-level mechanisms rather than composing existing primitives. For instance, transitioning a PD-disaggregated serving scheduler from min-TTFT~\cite{qin2025-mooncake} to a cache-affinity and load-balancing aware strategy~\cite{anonymous2025-dualmap} requires intrusive changes across the serving stack. This tight coupling restricts code reuse and limits the extensibility of tensor management strategies.

\nosection{Lack of programmable composition across tensor lifecycles.}
Beyond optimizing individual tensor management tasks, modern LLM services increasingly require composing multiple lifecycle primitives under workload-specific policies. For example, a distributed LLM serving system may need to simultaneously balance instance load, preserve KV cache locality, and migrate request states across instances \cite{zhang2026-lmetric}. Such optimization requires coordinating tensor placement, migration, and materialization with application-level decisions.
However, existing solutions expose these capabilities through fixed, task-specific interfaces. KV cache systems focus on cache storage, reuse, and transfer~\cite{kwon2023-pagedAttention, vllm-apc,
sglang-hicache, sglang-radixAttention, qin2025-mooncake}, while request schedulers
optimize routing decisions \cite{anonymous2025-dualmap}. Combining these mechanisms requires
intrusive integration across multiple components, preventing developers from flexibly implementing new tensor management policies. This limitation has also been identified from the perspective of LLM
execution engine design, where supporting emerging applications requires more programmable control over the inference process \cite{gim2025-pie}.
This limitation becomes increasingly important for emerging workloads with dynamic execution patterns. Reasoning approaches such as Chain-of-Thought (CoT)~\cite{wei2022-cot} and Tree-of-Thought (ToT)
~\cite{yao2023-tot} introduce intermediate execution states, such as reusable KV caches and branching context representations, that may need to be reused, duplicated, or migrated across instances. As such workloads become increasingly integrated into LLM applications and RL post-training rollouts~\cite{ouyang2022-rlhf, qin2025-seer, team2025-kimi1.5, wu2025-totrl}, supporting them requires a programmable tensor management abstraction rather than another task-specific optimization
stack.

\subsection{Requirements for Tensor-as-a-Service}\label{sec:tass-requirements}

The limitations above stem from the absence of a unified abstraction for managing tensor lifecycles across LLM workloads. Conceptually, modern LLM systems consist of tensor-producing and tensor-consuming
computation, while tensors increasingly become externally manageable states that require independent placement, movement, transformation, and reuse. However, existing systems embed these lifecycle mechanisms inside individual optimization stacks, limiting their composability and extensibility. Therefore, we propose decoupling tensor management from execution logic into an independent service layer that provides a state-management interface to execution engines as well as network and
storage backends (\figref{fig:tensor-management-layer}). This Tensor-as-a-Service (TaaS) paradigm must satisfy the following requirements:

\nosection{Tensor-native abstraction.}
The TaaS layer should natively understand tensor semantics and lifecycle, including different tensor states such as model weights and KV caches. Unlike traditional distributed object storage systems such
as Redis~\cite{redis}, Vineyard~\cite{yu2023-vineyard}, and Ray-Plasma ~\cite{ray-plasma}, which treat data as opaque blobs, a tensor management layer should maintain tensor identity, ownership, placement,
and transformation semantics. It should provide first-class handles for users to access and manipulate tensors while abstracting the underlying creation, transfer, storage, and transformation mechanisms through a high-performance implementation.

\nosection{Programmable lifecycle management.}
To support potentially infinite combinations of tensor management policies, the TaaS layer should provide programmable lifecycle primitives rather than hard-code optimization strategies. Different
from compute-centric frameworks such as Spark~\cite{zaharia2012-spark} and Ray~\cite{moritz2018-ray}, where users dispatch computation tasks, TaaS users orchestrate tensor states by composing lifecycle operations and defining workload-specific policies. This programmability enables system extensibility while maintaining a reusable tensor management substrate.

\nosection{Cluster-scale execution.}
As a unified service layer, TaaS must efficiently manage tensor states across cluster-scale resources with heterogeneous memory, network, and storage hierarchies. Achieving this requires distributed metadata management, fault tolerance, and a runtime environment with minimal overhead on execution paths.

\section{\sys Programming Framework}

To realize the TaaS paradigm described in \secref{sec:tass-requirements},
\sys exposes tensor lifecycle primitives through a programmable
framework that separates tensor management policies from execution
mechanisms. Specifically, \sys maps the lifecycle abstractions identified
in \secref{sec:lifecycle-primitives} into four programming abstractions:
\textit{Artifact} for tensor identity and ownership, \textit{Operation}
for tensor lifecycle manipulation, \textit{Plan} for composing
distributed workflows, and \textit{Signal} for policy feedback. This
section presents the programming model (\secref{sec:programming-model}),
the lifecycle APIs (\secref{sec:programming-apis}), and an illustrative
example (\secref{sec:programming-example}).

\subsection{Programming Model}\label{sec:programming-model}

\begin{figure}
	\centering
	\includegraphics[width=\linewidth]{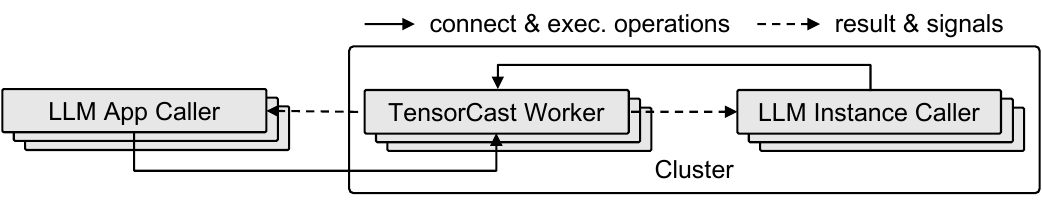}
	\caption{\sys's programming model.}
	\label{fig:programming-model}
\end{figure}

\sys separates tensor management policies from execution mechanisms
through a caller-worker programming model. Callers define workload-
specific tensor management strategies by composing lifecycle
operations, while \sys workers provide the distributed runtime that
stores tensors and executes these operations. As \figref{fig:programming-model}
illustrates, users write tensor management programs on \textit{callers};
each caller connects to a \sys worker to obtain a \texttt{Runtime} handle.
Workers maintain tensor states, execute lifecycle operations in a
distributed manner, and return execution results together with internal
state signals to callers for error handling, tracing, and policy
decisions. Based on their roles, callers fall into two categories:

\nosection{Application Caller.}
Application callers implement workload-level tensor management policies.
They may physically reside within a \sys cluster but remain logically
external to the tensor management runtime, accessing \sys as a service
for LLM applications. Typical examples include request routers,
auto-scaling managers, and orchestrators for multi-instance inference
(\eg ToT reasoning). These callers explicitly coordinate tensor
operations with application-level computation (\eg prefill and decode),
enabling fine-grained control over tensor lifecycles without modifying
the underlying execution engines. We later demonstrate an application
caller through a programmable KV cache rebalancing policy in
\secref{sec:programming-example}.

\nosection{Instance Caller.}
Instance callers provide the mechanism boundary between \sys and
execution engines. They operate logically inside the \sys cluster and
typically integrate directly with inference or training instances to
expose engine-resident tensor states. For example, to support KV cache
sharing, a minimal SGLang integration caller implements the HiCache
\cite{sglang-hicache} interfaces while connecting to a \sys worker,
allowing the entire \sys cluster to serve as the KV backend. Unlike
existing task-specific integrations such as Mooncake
\cite{qin2025-mooncake} and LMCache \cite{liu2025-lmcache}, which
typically couple tensor movement mechanisms with optimization policies,
\sys's instance integration fixes only the mechanism required to access engine-resident tensors. The management policy remains programmable through callers, allowing different tensor management strategies to be implemented without modifying the execution engines.

\begin{table*}[htbp]
    \centering
    \begin{tabularx}{\textwidth}{l l X}
        \toprule
        \textbf{Category} & \textbf{Class \& Function} & \textbf{Description} \\
        \midrule

        \textit{Artifact}
         & \texttt{register(tensor, art\_id) -> Artifact} & Register a caller-leased tensor to \sys. \\
         & \texttt{put(tensor, art\_id) -> Artifact} & Put a system-owned tensor to \sys. \\
         & \texttt{artifact(art\_id) -> Artifact} & Get an \texttt{Artifact} handle from \sys given an ID. \\
         & \texttt{Artifact.tensor\_meta() -> TensorMeta} & Get the tensor metadata of an \texttt{Artifact}. \\
         & \texttt{Artifact.tensor\_dict() -> Tensor} & Materialize a tensor from an \texttt{Artifact}. Blocking. \\
         & \texttt{Artifact.view(slice, name) -> Artifact} & Derive a new \texttt{Artifact} from a view of an existing one. \\
        \midrule
        
        \textit{Plan} 
         & \texttt{context(id, ddl, key) -> CallContext} & Create a new call context. \\
         & \texttt{plan(ctx) -> Plan} & Create an executable plan with a context. \\
         & \texttt{Plan.on\_worker(worker) -> PlanStepBuilder} & Build a new plan step on a \sys worker. \\
         & \texttt{Plan.on\_instance(inst) -> PlanStepBuilder} & Build a new plan step on an instance. \\
         & \texttt{Plan.run(concurrency) -> PlanResult} & Execute the plan with specified concurrency. Blocking. \\
        \midrule

        \textit{Worker}
         & \texttt{prefetch(artifact, device) -> PlanStep} & Transfer an \texttt{Artifact} replica to the worker on \texttt{device}. \\
        \textit{Operation}
         & \texttt{prefetch\_many(artifacts, device) -> PlanStep} & Batched \texttt{prefetch} call. \\
         & \texttt{pin(artifact, device) -> PlanStep} & Pin an \texttt{Artifact} replica on the worker's \texttt{device}. \\
        \midrule

        \textit{Instance}
         & \texttt{publish(req\_id, ttl) -> PlanStep} & Flush a request's KV cache to \sys. \\
        \textit{Operation}
         & \texttt{manifest(req\_id) -> PlanStep} & Get the set of \texttt{Artifact}s of a request's KV cache. \\
         & \texttt{hydrate(req\_id, artifacts) -> PlanStep} & Load a request's KV caches into the instance. \\
         & \texttt{evict(req\_id) -> PlanStep} & Evict a request's KV caches on the instance. \\
         & \texttt{transform\_into(artifact, spec, target)} & Transform the \texttt{artifact} tensor according to \texttt{spec} rule \\
         & \texttt{-> PlanStep} & and fill into the instance-owned \texttt{target} buffer. \\
        \midrule
        
        \textit{Signal} 
         & \texttt{connect(worker\_addr) -> Runtime} & Connect to a worker and get a runtime. \\
         & \texttt{list\_instances() -> list[Instance]} & List the statuses of all instances. \\
         & \texttt{list\_workers() -> list[Worker]} & List the statuses of all \sys workers. \\
        
        \bottomrule
    \end{tabularx}
    \caption{Overview of essential \sys programming APIs.}
    \label{tab:apis}
\end{table*}

\subsection{Lifecycle Programming Abstractions}\label{sec:programming-apis}

\sys provides four programming abstractions that realize the tensor
lifecycle primitives described in \secref{sec:lifecycle-primitives}.
\textit{Artifact} represents tensor states and ownership;
\textit{Operation} provides lifecycle manipulation primitives;
\textit{Plan} composes distributed tensor workflows; and
\textit{Signal} exposes runtime information required for policy
decisions. Together, these abstractions separate tensor management
policies from execution mechanisms while providing a unified interface
for diverse LLM workloads. \tabref{tab:apis} summarizes the essential
APIs.

\nosection{Artifact.}
It is the first-class representation of tensor states in
\sys. It decouples tensor identity, ownership, and lifecycle from the
underlying physical representation, allowing callers to manipulate
tensors without explicitly managing their storage locations or device
layouts.
From the caller's perspective, an \texttt{Artifact} is a logical handle
to a tensor stored in the unified tensor pool. The actual tensor content
is maintained by \sys workers as distributed replicas and is materialized
only when callers request a concrete tensor representation through
\texttt{tensor()}. This separation enables \sys to independently
optimize tensor placement, replication, and movement while preserving a
stable interface for callers. 
\texttt{Artifact}s are immutable and maintain a canonical identity,
including tensor metadata, format, and content information, to
deterministically identify tensor states. Callers introduce tensors into
\sys under two ownership models: 
\textit{i) System-Owned.}
\sys assumes responsibility for the tensor lifecycle, which is suitable
for durable global tensors such as model weights;
\textit{ii) Caller-Leased.}
Callers retain ownership while granting \sys temporary management
authority through a lease. This model supports transient but shareable
tensors, such as KV caches during migration.

\nosection{Operation.}
Unlike general-purpose distributed frameworks that execute arbitrary user
code (\eg Ray~\cite{moritz2018-ray}), \sys exposes a finite but
extensible set of tensor lifecycle operations. This design provides a
stable optimization boundary: callers define management policies by
composing operations, while \sys controls their distributed execution.
An operation carries a \texttt{CallContext} and behaves as a lazy handle
that blocks only when callers explicitly await its result. The context
contains a tracing ID, a deadline for maximum execution duration, and an
idempotency key for reentrant retrying. Based on the execution location,
operations are categorized into two types:
\textit{i) Worker Operation.}
These operations operate entirely within the TaaS layer without
interacting with computational logic. They manipulate tensor states
inside \sys, such as processing tensors locally on a worker
(\eg \texttt{pin()}) or transferring replicas between workers
(\eg \texttt{prefetch()});
\textit{ii) Instance Operation.}
These operations interact with computation engines and manipulate
engine-resident tensor states through instance callers. Representative
examples include publishing a request's KV cache into the \sys cluster
as a set of \texttt{Artifact}s through \texttt{publish()} and loading
them back through \texttt{hydrate()}, as well as transforming model
weights and filling rank-specific VRAM parameter buffers through
\texttt{transform\_into()}. The instance adaptor residing within the
execution instance implements these operations.

\nosection{Plan.}
It provides the composition abstraction for coordinating
multiple lifecycle operations across distributed workers and instances.
Each operation in a plan corresponds to a \texttt{PlanStep}, and \sys
maintains step dependencies through a Directed Acyclic Graph (DAG).
Callers initiate execution through \texttt{run()}, after which \sys
executes operations according to the dependency graph and specified
concurrency level.
All operations within a plan share the same execution context.
Consequently, \sys traces and retries a plan as an idempotent workflow
using a unified ID and idempotency key. \texttt{Plan} allows callers to
orchestrate complex tensor workflows without manually managing operation
ordering, synchronization, or failure handling.
\sys deliberately avoids distributed transaction semantics. A plan
provides dependency ordering, tracing, deadline propagation, and
idempotent retry, but does not provide ACID atomicity or rollback.
Partially executed plans --- such as a published KV cache or a prefetched
replica --- may remain and are reused, retried, or cleaned up according
to operation semantics. This design matches LLM tensor lifecycles,
which predominantly create immutable replicas, materialize tensor views,
or move recomputable states without requiring general-purpose distributed transactions.

\nosection{Signal.}
It closes the policy loop by exposing runtime statuses that
callers need to make tensor management decisions. Instead of embedding
placement or scheduling policies inside \sys, the system provides
observations that enable callers to implement adaptive management
strategies.
Prominent signals include \texttt{list\_workers()}, which returns the
status of individual workers such as memory and network pressure, and
\texttt{list\_instances()}, which returns the status of individual
instances such as load level and the corresponding worker. By combining
these runtime signals with the \texttt{Plan} abstraction, callers can
formulate and execute diverse tensor management policies while reusing
the same underlying lifecycle mechanisms.

\subsection{Example: Programmable KV Cache Rebalancing}
\label{sec:programming-example}

\lstref{code:tc-router} demonstrates how an application caller can
implement a workload-specific tensor management policy by composing
\sys lifecycle abstractions. The example implements a programmable KV
cache rebalancing policy for distributed LLM serving, which jointly
considers instance load and KV cache locality without modifying the
underlying inference engine or KV cache backend.

The caller first uses \texttt{Signal} APIs to observe the runtime states
of serving instances and applies an external policy to select a request,
together with the KV cache of its entire context, for migration between
instances (lines 9--10). The caller then constructs a \texttt{Plan} with
a \texttt{CallContext} to compose three lifecycle operations:
(1) an instance operation \texttt{publish()} flushes the KV cache from
the source instance into the \sys cluster and registers the
corresponding \texttt{Artifact}s;
(2) a worker operation \texttt{prefetch\_many()} optionally transfers
the artifacts to the worker connected to the destination instance; and
(3) an instance operation \texttt{hydrate()} materializes the KV cache
inside the destination inference engine. The prefetch step is optional
because \texttt{hydrate()} can automatically fetch required artifacts
once they are registered in \sys.
This example highlights the programmability enabled by \sys: callers
control tensor management policies, while \sys provides reusable
mechanisms for tensor storage, movement, and materialization. The same
programming model can be extended to other workload-specific tensor
management strategies. We evaluate this lightweight caller-driven rebalancing
policy in \secref{sec:eval-router}.

\begin{lstlisting}[caption={LLM router with request-level KV cache rebalancing with \sys APIs. Error-handling codes are emitted for simplicity.}, label={code:tc-router}]
import tensorcast as tc

# Initialize the caller's TensorCast runtime
rt = tc.connect(gateway_addr) 

def rebalance():
  # Choose to rebalance a request from inst1
  # to inst2 decided by a rebalancing policy
  instances = rt.signals().list_instances()
  inst1, inst2, req_id = decide(instances)

  # Perform KV cache migration with a Plan
  ctx = tc.CallContext(id="tracing_id", deadline_ms=5000, idempotency_key="idem_key")
  plan = tc.Plan(ctx)
  # Flush request's KV cache as artifacts 
  # from inst1 into the cluster
  flush_res = plan.on_instance(inst1).publish(req_id)
  # Optionally prewarm all KV cache 
  # on the worker CPU DRAM connected by inst2 
  plan.on_worker(inst2.worker).prefetch_many(flush_res.artifact_result, device="cpu")
  # Load request's KV cache into inst2
  plan.on_instance(inst2).hydrate(req_id, flush_res.artifact_result)
  # Execute the composed lifecycle workflow
  plan.run(concurrency=1)
\end{lstlisting}

\section{System Architecture}\label{sec:sys-arch}

\begin{figure}
	\centering
	\includegraphics[width=\linewidth]{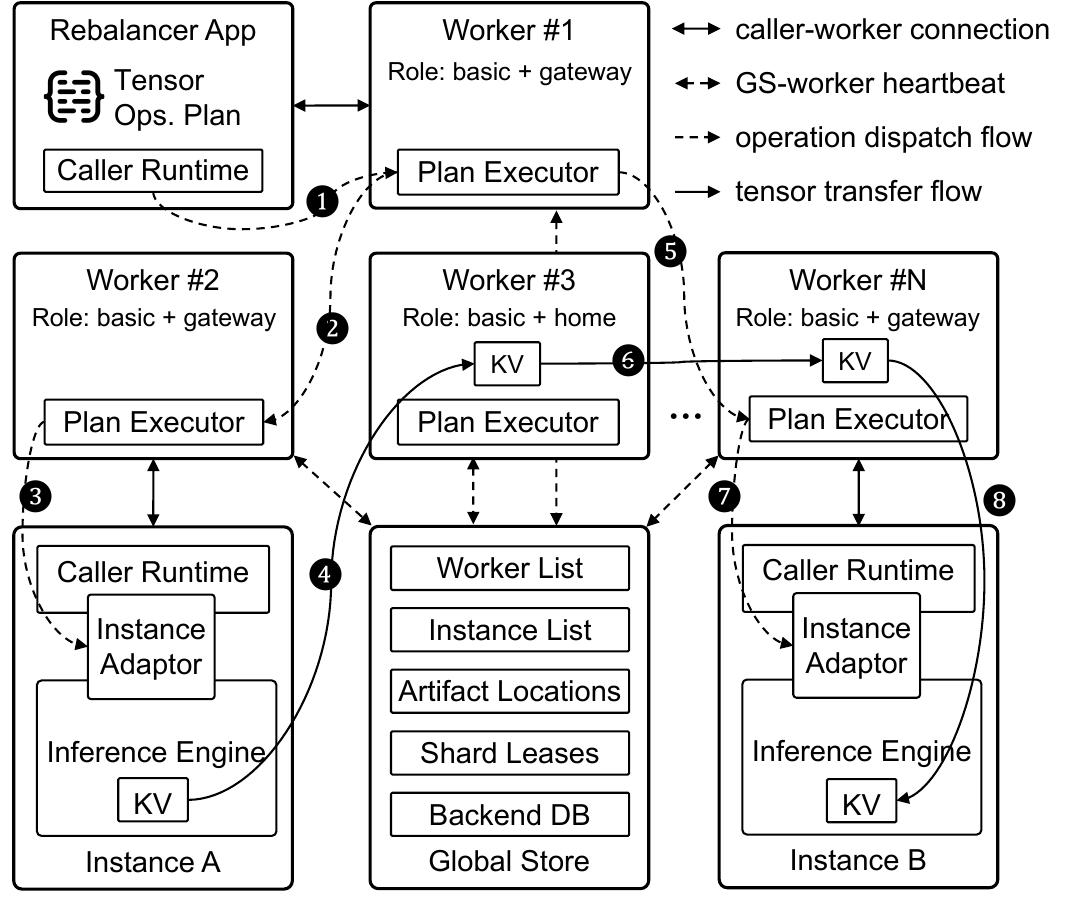}
	\caption{\sys system architecture. All components belong to the same \sys cluster except for the \textit{Req. Router App} process. Steps \textnormal{\protect\WoB{1}--\protect\WoB{8}} illustrate the workflow of executing the KV migration task shown in \lstref{code:tc-router} (line 12--23). See detailed explanation in \secref{sec:arch-illustration}.}
	\label{fig:sys-arch}
\end{figure}

\subsection{System Components}

The \sys runtime realizes the caller-worker programming model described
in \secref{sec:programming-model}. As shown in
\figref{fig:sys-arch}, a \sys cluster consists of worker nodes,
instance nodes, and a lightweight global store (GS). Callers specify
tensor management policies through \sys APIs, while workers execute
lifecycle operations over distributed tensor states. Instance nodes
provide the mechanism boundary between \sys and computation engines,
allowing \sys to manage engine-resident tensors without embedding
workload-specific policies into the execution layer.

\nosection{Worker.}
A worker is the fundamental execution unit of \sys, responsible for
maintaining tensor states and executing lifecycle operations. Rather
than deploying separate services for storage, workflow execution, and
metadata management, \sys composes worker functionalities through
different \textit{roles}. Currently, three roles exist:
\textit{i) Basic.}
A worker with the basic role contributes its memory and disk resources
to the unified tensor pool (\secref{sec:arch-tensor-pool}). It also
maintains a local plan executor for processing worker-local plan steps
(\eg artifact \texttt{prefetch()}) dispatched from other workers, but
does \textit{not} directly receive plans from callers. All \sys workers
operate with the basic role at minimum.
\textit{ii) Gateway.}
Gateway workers provide the entry point for callers to submit lifecycle
workflows. They maintain caller connections, resolve plan-step
dependencies, and dispatch operations to corresponding workers or
instances for execution. In practice, gateway workers can share a
common routable address behind a load balancer to provide scalable
access to external callers.
\textit{iii) Shard Home.}
For high-cardinality tensors such as KV caches, \sys partitions tensor
states into multiple \textit{shards} (\secref{sec:arch-tensor-pool}).
Workers with the shard home role maintain the ownership and consistency
invariants of specific shards. 
This composable role architecture separates different lifecycle
responsibilities while preserving a unified worker abstraction. It
simplifies configuration, deployment, and monitoring of heterogeneous worker resources (\eg assigning high-performance nodes as KV cache shard homes) and
allows \sys to introduce new lifecycle capabilities through additional
worker roles.

\nosection{Instance.}
Instance nodes implement the mechanism boundary between \sys and
computation engines. A typical instance node runs an LLM execution
engine (\eg an inference or training framework) and contains a \sys
caller runtime that connects to a gateway worker. \sys integrates with
the execution engine through an \textit{Instance Adaptor}, which
implements the minimal interfaces required to export, import, or
transform engine-resident tensor states.
This separation ensures that instance integration fixes only the
underlying tensor management mechanisms, while workload-specific
policies remain programmable through callers. Consequently, new tensor
management strategies can evolve without repeatedly modifying execution
engines or backend implementations.

\nosection{Global Store.}
\sys uses a Global Store (GS) as a lightweight control-plane metadata
service rather than a component in the tensor data path. The GS
maintains cluster metadata, including worker and instance status, the
replica locations of low-cardinality tensors
(\secref{sec:arch-tensor-pool}), and shard ownership information.
To avoid introducing a centralized bottleneck, workers cache metadata
locally and refresh it through update exchanges with the GS. Therefore,
most tensor management workflows bypass the GS and execute directly
among workers and instances. The GS additionally serves as the entry
point for cluster observability, including workload monitoring and
runtime metrics.

\subsection{Unified Tensor Pool}\label{sec:arch-tensor-pool}

The unified tensor pool realizes the \texttt{Artifact} abstraction by
separating logical tensor identity from physical tensor residency.
Although callers interact with stable \texttt{Artifact} handles,
\sys dynamically manages concrete tensor replicas across workers,
devices, and storage tiers. Since different tensor states exhibit
different lifecycle characteristics, \sys adopts different management
policies for \textit{low-cardinality} and \textit{high-cardinality} tensors.

\nosection{Low-Cardinality Tensors.}
Some tensors in LLM workloads are immutable, durable, and accessed
relatively infrequently. Representative examples include static model
weights, which typically have a small number of tensor identities but
large data sizes. We classify such tensors as low-cardinality tensors. Since their metadata changes infrequently, \sys maintains their
replica locations centrally through the GS.
During operations involving low-cardinality tensors, workers query the
GS to locate available replicas. For example, an inference instance may
load a Qwen3.5 model from local storage and introduce the weights into
the tensor pool through
\texttt{put(weights, "Qwen/Qwen3.5-35B-A3B")}. \sys creates an
\texttt{Artifact} replica on a worker, or directly generates a CUDA IPC
handle when the worker and instance reside on the same host. Subsequent
instances initializing the same model can obtain the corresponding
\texttt{Artifact} and materialize the weights through
\texttt{tensor()}, fetching replicas from the original worker or nearby
workers populated through prior \texttt{prefetch()} operations.
In practice, a newly launched inference instance is typically deployed
together with a co-host or nearby \sys gateway worker, allowing model weights to be shared with CUDA IPC or transferred through GPU Direct RDMA. Because low-cardinality tensors contain few distinct
metadata entries and incur infrequent metadata accesses, centralized
metadata management through the GS does not become a scalability
bottleneck.

\nosection{High-Cardinality Tensors.}
In contrast, other tensor states exhibit high cardinality: they contain
many fragmented objects, are frequently created and reused, and
require dynamic placement decisions. KV caches are a representative
example. Inference engines continuously generate and consume KV cache
blocks or pages, making them significantly more dynamic than model weights.
Maintaining the locations of all such tensors directly in the GS would
introduce excessive metadata traffic.
To address this challenge, \sys partitions high-cardinality tensors into
\textit{shards} and assigns each shard to a worker with the shard home
role. The shard home worker maintains the ownership and consistency
invariants of its shard. To prevent split-brain ownership during worker
failures or network partitions, \sys associates each shard with a
\textit{shard lease} containing the shard home worker, expiration time,
and a monotonically increasing fencing token. Assume $N$ shards exist. For a given $\mathtt{shard\_id} \in \{0, 1, \dots, N-1\}$, workers with the shard
home role compute their ranks using Highest Random Weight \cite{thaler2002-hrw} hashing:
$\mathtt{rank} = \mathrm{HRW}(\mathtt{shard\_id}, \mathtt{worker\_id})$.
The top-$k$ workers for each shard (we use $k=3$ in our implementation
to guarantee lease acquisition and reduce contention) compete to acquire
the corresponding shard lease through the GS. The GS grants the lease
to the first arriving worker and remains responsible only for lease
metadata, while workers cache lease records locally.
Shard home workers periodically renew leases through heartbeat
messages. During worker or network failures, an expired lease triggers
a new ownership acquisition process. After a new shard home worker
acquires the lease, the fencing token increases, invalidating operations
from stale workers because all workers reject high-cardinality tensor
operations carrying outdated fencing tokens. This lease-based design
distributes tensor metadata management across workers and limits GS
operations to shard number $N$, rather than per-tensor queries,
allowing \sys to scale to high-frequency tensor state queries.

\subsection{Distributed Plan Execution}\label{sec:arch-plan-exec}

\sys executes the \texttt{Plan} abstraction in a distributed manner,
allowing callers to compose tensor lifecycle workflows without
introducing a centralized execution bottleneck. Rather than maintaining
a global operation scheduler, \sys distributes plan execution across
workers, where each worker locally executes lifecycle operations assigned to it. A single operation is represented as a one-step plan, providing a unified execution path for all tensor workflows.

\nosection{Worker-Local Plan Execution.}
Callers submit plans to a connected gateway worker and do not manage
distributed execution logic locally. The gateway worker coordinates plan
execution by resolving step dependencies and dispatching individual
operations to corresponding workers or instances. This design keeps
caller programs lightweight while maintaining a clean boundary between \sys system and the caller SDK.
Each worker maintains a local plan executor that supports three execution
responsibilities:
\textit{i)}
Execute worker-local operations, such as pinning tensor replica
residency or transferring replicas between workers through
\texttt{prefetch()};
\textit{ii)}
Invoke instance adaptors to execute operations that interact with
engine-resident tensor states, such as publishing or hydrating KV caches;
\textit{iii)}
Maintain plan execution progress, including dependency resolution,
operation dispatch, retry handling, and result propagation. Gateway
workers provide this coordination functionality for caller-submitted
plans.
At the control plane, workers discover each other through worker
metadata cached from the GS and communicate through RPCs. At the data
plane, tensor contents bypass the GS and are transferred directly
between workers and instances through high-performance communication
paths, such as P2P RDMA when available.

\subsection{Lifecycle Consistency and Fault Tolerance}

\sys adopts fault tolerance mechanisms that preserve tensor lifecycle
semantics under failures. Different tensor states follow different
recovery strategies: durable low-cardinality tensors rely on persistent
replicas, while high-cardinality tensors rely on distributed shard
ownership and lease-based consistency.

\nosection{Worker Failover.}
Workers maintain cached execution metadata, while persistent metadata
remains recoverable from the GS. Following a worker failure, recovered
workers reconstruct transient caches, such as worker and instance lists,
and reacquire shard leases from the GS. Low-cardinality tensor replicas
remain available if their physical replicas survive or reside on
persistent storage. In contrast, high-cardinality tensors whose shard
leases expire are considered unavailable until they are reconstructed or
republished.

\nosection{Global Store Failover.}
The GS maintains persistent metadata required for cluster coordination,
including worker status, replica information, and shard ownership. This
metadata is committed to a backend database for durability. In
production deployments, standard replication mechanisms such as state
machine replication (\eg Paxos~\cite{lamport2001-paxos}, Raft
~\cite{ongaro2014-raft}) or chain replication
~\cite{van2004-chain, terrace2009-craq} can provide GS availability.

\subsection{Putting Everything Together}\label{sec:arch-illustration}

We use the programmable KV cache rebalancing example from
\lstref{code:tc-router} (lines 12--23) to illustrate how \sys realizes
a complete tensor lifecycle workflow. As \figref{fig:sys-arch}
illustrates, the rebalancer application, instance A, and instance B
connect to workers 1, 2, and $N$, respectively. When the application
caller invokes \texttt{plan.run()}, the plan is submitted to worker 1,
which acts as the gateway worker responsible for coordinating execution
(\WoB{1}).
To execute the \texttt{publish()} operation on instance A, worker 1
dispatches the corresponding plan step to worker 2, which is connected
to instance A (\WoB{2}). The plan executor on worker 2 invokes the
instance adaptor inside instance A to export the engine-resident KV
cache (\WoB{3}). The instance adaptor retrieves the KV cache from the
inference engine, determines the corresponding shard, locates the shard
home worker 3 using the locally cached shard lease, and transfers the
KV cache to worker 3 (\WoB{4}).
After the KV cache is registered as \texttt{Artifact}s in \sys, worker 1
dispatches the optional \texttt{prefetch\_many()} operation to worker
$N$ (\WoB{5}). Worker $N$ uses its local lease information to locate
the shard home worker 3 and transfers the required KV replicas locally
(\WoB{6}). Finally, worker $N$ invokes the instance adaptor inside
instance B (\WoB{7}) to execute \texttt{hydrate()}, which materializes
the KV cache back into the destination inference engine (\WoB{8}).
The entire data path is distributed across workers and instances without GS, and the GS remains responsible only for control-plane metadata such as shard ownership and leases.

\section{Implementation}\label{sec:impl}

We implement \sys as a distributed C++ runtime with a Python SDK.
The runtime consists of approximately 160K lines of C++ code, while
the SDK contains approximately 95K lines of Python code. We highlight
key implementation details of tensor materialization, data transfer,
and LLM framework integration below. The implementation is publicly available at \url{https://github.com/tensorcast-ai/tensorcast}.

\nosection{Runtime Implementation.}
The \sys runtime is implemented as a distributed C++ service.
Workers communicate through gRPC and maintain tensor replicas across
CPU memory, GPU memory, and local storage. The global store is
implemented as a Python gRPC service backed by DuckDB
\cite{duckdb}, maintaining control-plane metadata including worker
status, artifact indices, replica locations, and shard leases.
Workers do not access the global store during tensor data transfer.
Instead, they cache required metadata locally and execute operations
through the distributed plan runtime. Worker liveness and metadata
consistency are maintained through heartbeat and reconciliation RPCs.

\nosection{Tensor Materialization and Transfer.}
To realize Artifact materialization efficiently, workers implement a
streaming pipeline that resolves tensor metadata, allocates target
memory, transfers tensor contents from local storage or remote workers,
verifies integrity, and exports device-specific handles to callers.
GPU-resident tensors on the same host are exposed through CUDA IPC to
avoid unnecessary data copies.
For inter-worker tensor movement, \sys uses RDMA when possible and falls back to a multi-connection TCP
transport based on userspace mTCP~\cite{jeong2014-mtcp}. GPU transfers
use pinned streaming buffers and asynchronous CUDA streams to overlap data movement with execution.

\nosection{Programming Interface.}
The Python SDK provides the programming interface used by application
callers and instance callers. It exposes TensorCast abstractions through
a PyTorch-compatible interface and communicates with local workers
through protobuf-generated gRPC stubs. The SDK manages caller-side
runtime state, including session context, deadlines, retries, metadata
caching, and the lifecycle of returned shared-memory handles.
The SDK deliberately does not communicate directly with the global
store, preserving the separation between application programs and the
TensorCast control plane.

\begin{figure*}
	\centering
	\begin{subfigure}[t]{0.49\linewidth}
		\captionsetup{margin={4pt,0pt}}
		\includegraphics[width=\linewidth]{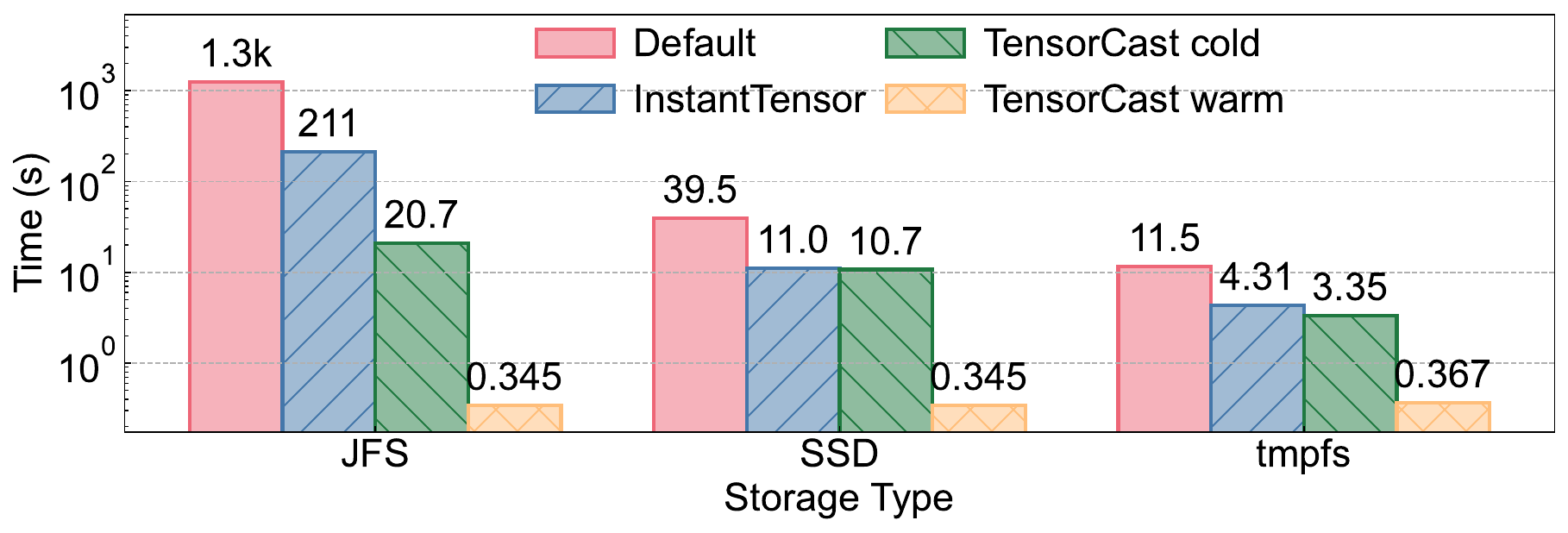}
		\caption{Weight load time.}
		\label{fig:eval-loader-30b-data_plane}
	\end{subfigure}
    \hfill
	\begin{subfigure}[t]{0.49\linewidth}
		\captionsetup{margin={0pt,4pt}}
		\includegraphics[width=\linewidth]{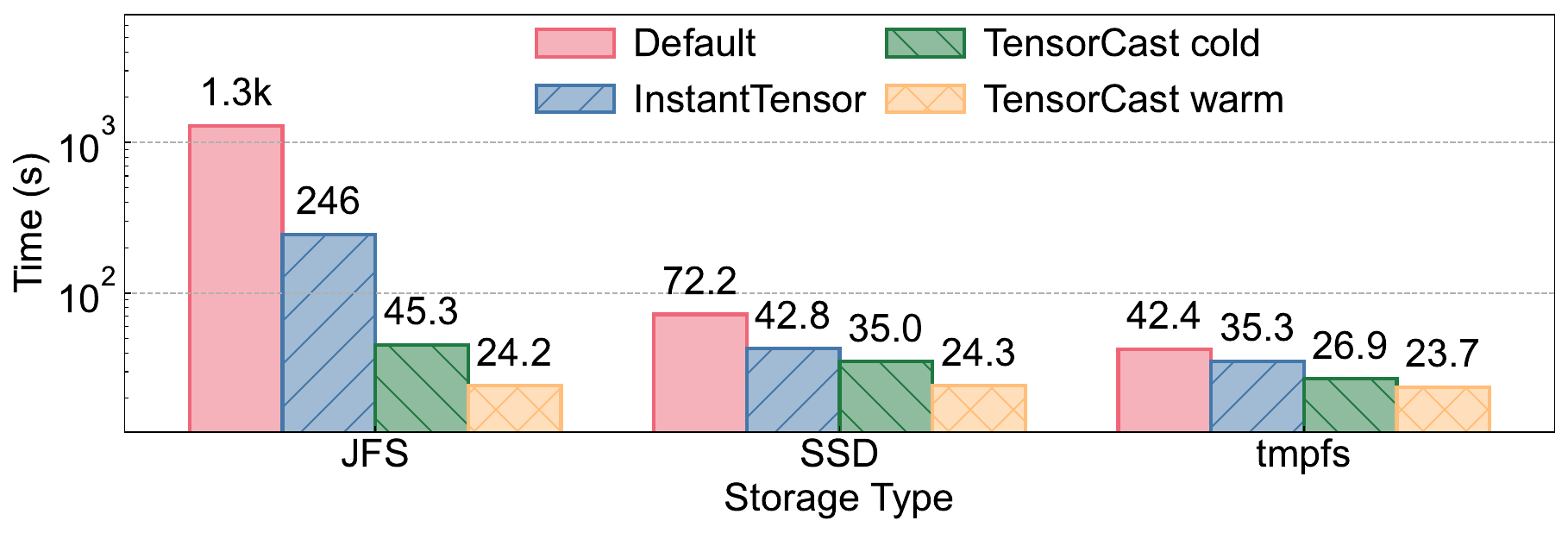}
		\caption{End-to-end ready time.}
		\label{fig:eval-loader-30b-e2e}
	\end{subfigure}
    \caption{Instance launch time for Qwen3-30B-A3B model.}
    \label{fig:eval-loader-30b}
\end{figure*}

\begin{figure*}
	\centering
	\begin{subfigure}[t]{0.49\linewidth}
		\captionsetup{margin={4pt,0pt}}
		\includegraphics[width=\linewidth]{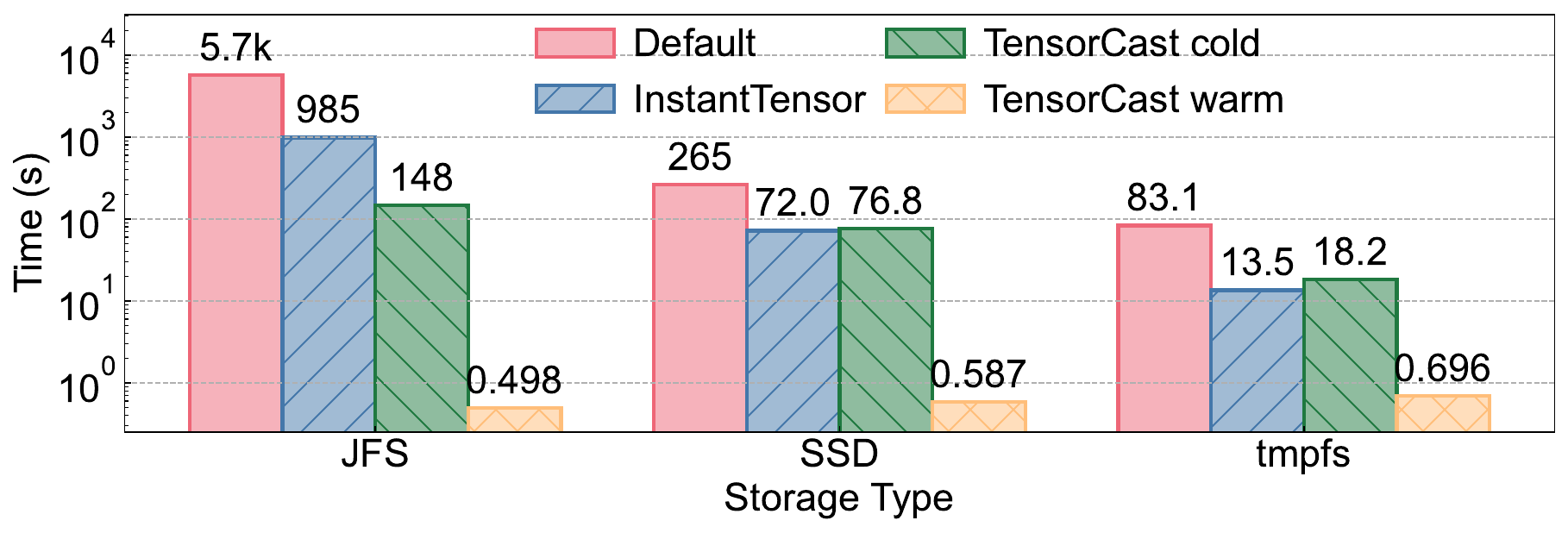}
		\caption{Weight load time.}
		\label{fig:eval-loader-235b-data_plane}
	\end{subfigure}
    \hfill
	\begin{subfigure}[t]{0.49\linewidth}
		\captionsetup{margin={0pt,4pt}}
		\includegraphics[width=\linewidth]{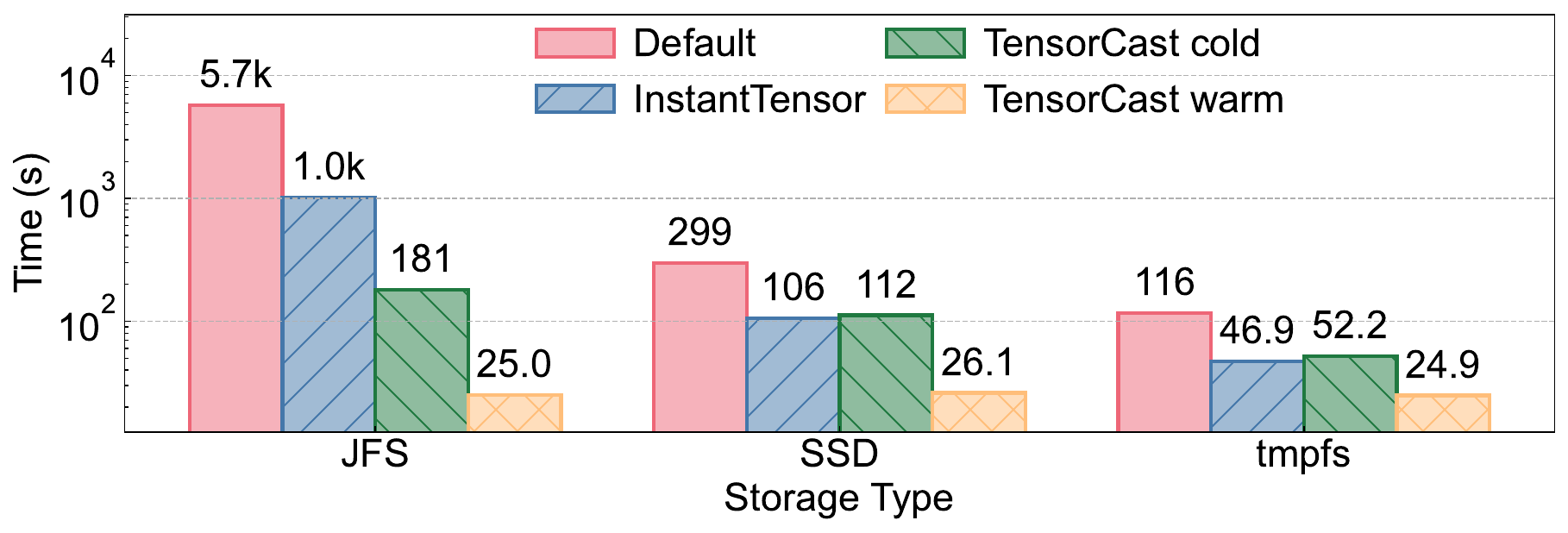}
		\caption{End-to-end ready time.}
		\label{fig:eval-loader-235b-e2e}
	\end{subfigure}
    \caption{Instance launch time for Qwen3-235B-A22B model.}
    \label{fig:eval-loader-235b}
\end{figure*}

\section{\sys Showcases \& Evaluation}\label{sec:eval}
We evaluate \sys to answer two questions:
\textit{i)} whether a unified tensor lifecycle layer can provide
efficient mechanisms comparable to specialized tensor management
systems; and
\textit{ii)} whether its programmable abstractions enable new tensor
management policies through composition of lifecycle primitives.
We integrate \sys with mainstream inference engines, vLLM
\cite{vllm} and SGLang \cite{zheng2024sglang}, and evaluate four
representative scenarios:
model weight materialization (\secref{sec:eval-loader}),
model weight synchronization (\secref{sec:eval-updater}),
KV cache management (\secref{sec:eval-kv}), and programmable request
routing (\secref{sec:eval-router}).
The first three scenarios evaluate the efficiency of \sys's reusable
tensor lifecycle mechanisms, while the last demonstrates the
programmability enabled by composing these mechanisms into a
workload-specific policy.

\begin{figure}
	\centering
	\begin{subfigure}[t]{0.48\linewidth}
		\captionsetup{margin={0pt,4pt}}
		\includegraphics[width=\linewidth]{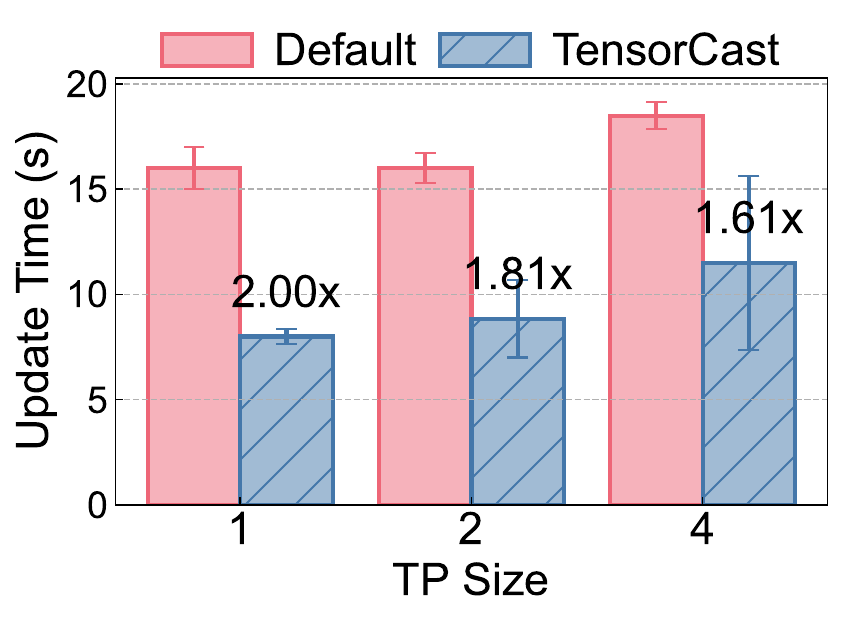}
		\caption{Qwen3-14B.}
		\label{fig:eval-updater-14b}
	\end{subfigure}
	\hfill
	\begin{subfigure}[t]{0.48\linewidth}
		\captionsetup{margin={4pt,0pt}}
		\includegraphics[width=\linewidth]{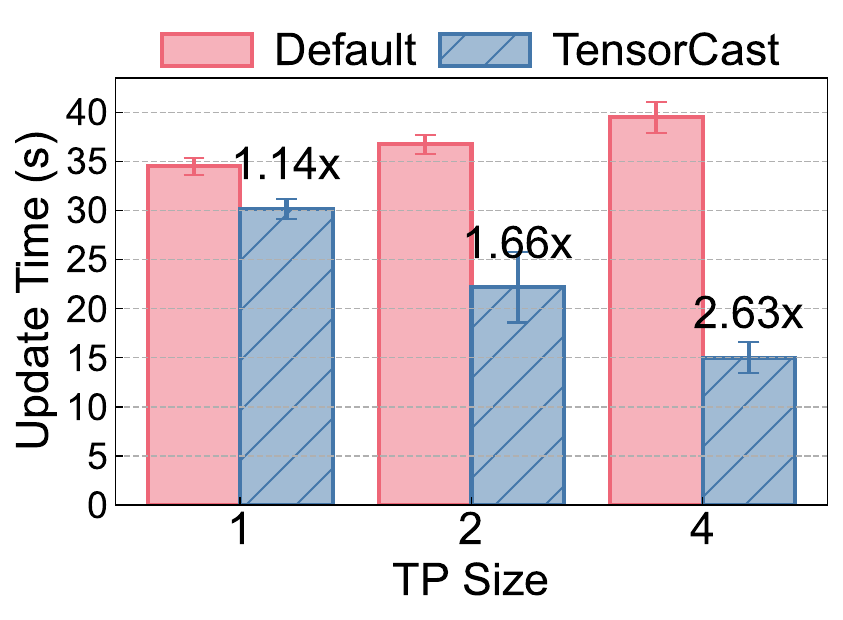}
		\caption{Qwen3-32B.}
		\label{fig:eval-updater-32b}
	\end{subfigure}
    \caption{Model weight synchronization time.}
    \label{fig:eval-updater}
\end{figure}

\begin{figure*}[t]
	\centering
	\begin{subfigure}[t]{0.24\linewidth}
        \centering
		\includegraphics[width=\linewidth]{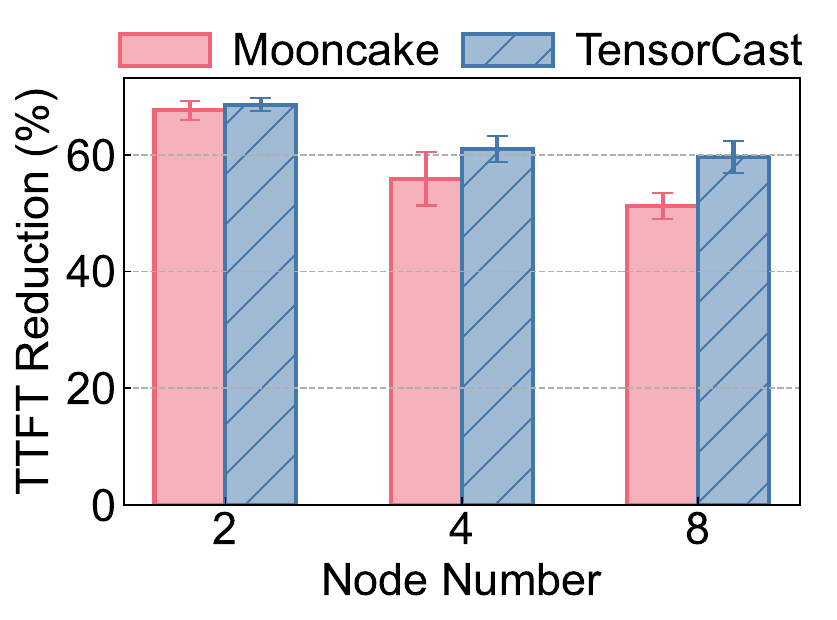}
		\caption{16k prompt w/ RDMA.}
		\label{fig:eval-kv-32b-16k-rdma}
	\end{subfigure}
	\hfill
	\begin{subfigure}[t]{0.24\linewidth}
        \centering
		\includegraphics[width=\linewidth]{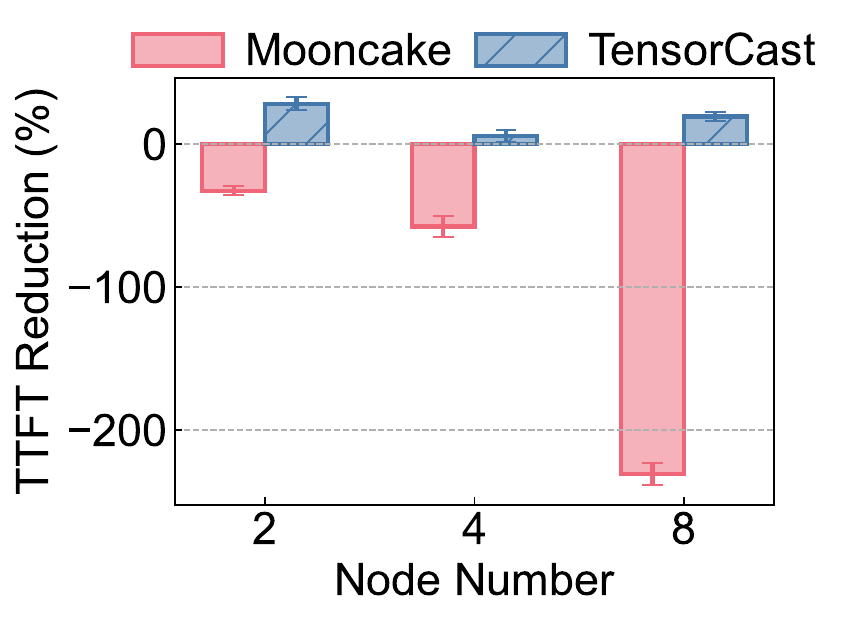}
		\caption{16k prompt w/o RDMA.}
		\label{fig:eval-kv-32b-16k-tcp}
	\end{subfigure}
    \hfill
	\begin{subfigure}[t]{0.24\linewidth}
        \centering
		\includegraphics[width=\linewidth]{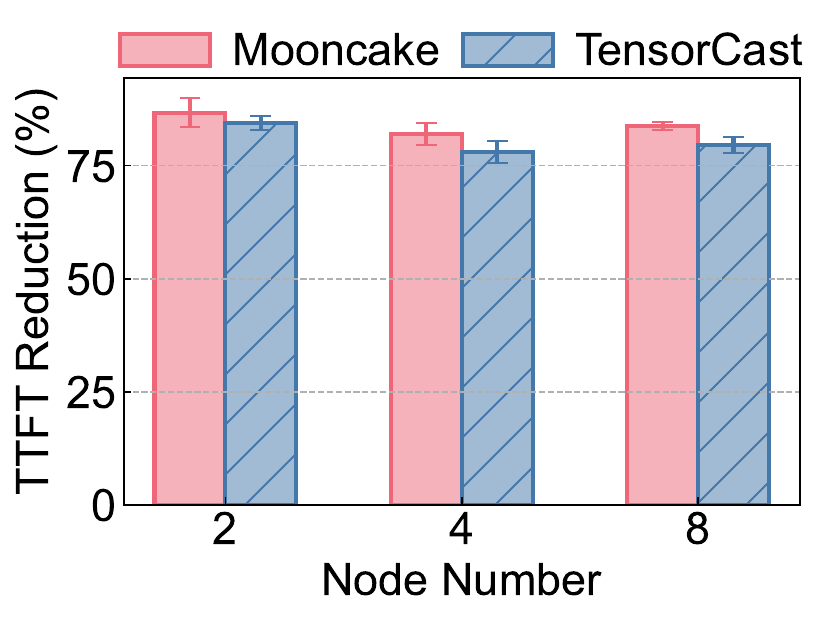}
		\caption{35k prompt w/ RDMA.}
		\label{fig:eval-kv-32b-35k-rdma}
	\end{subfigure}
	\hfill
	\begin{subfigure}[t]{0.24\linewidth}
        \centering
		\includegraphics[width=\linewidth]{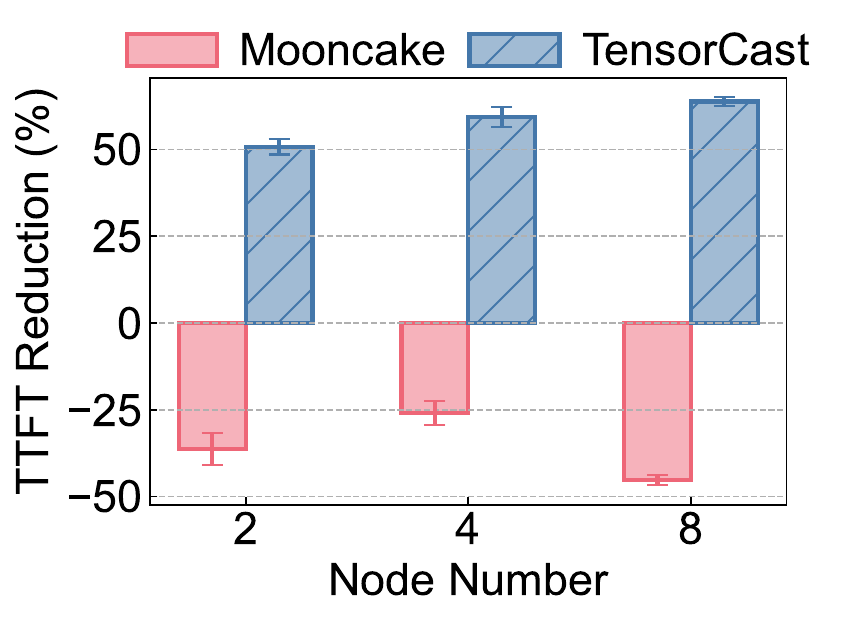}
		\caption{35k prompt w/o RDMA.}
		\label{fig:eval-kv-32b-35k-tcp}
	\end{subfigure}
    \caption{Qwen3-32B TTFT redcution ratio with \sys and Mooncake KV storage backend.}
    \label{fig:eval-kv-32b}
\end{figure*}

\begin{figure*}[t]
	\centering
	\begin{subfigure}[t]{0.24\linewidth}
        \centering
		\includegraphics[width=\linewidth]{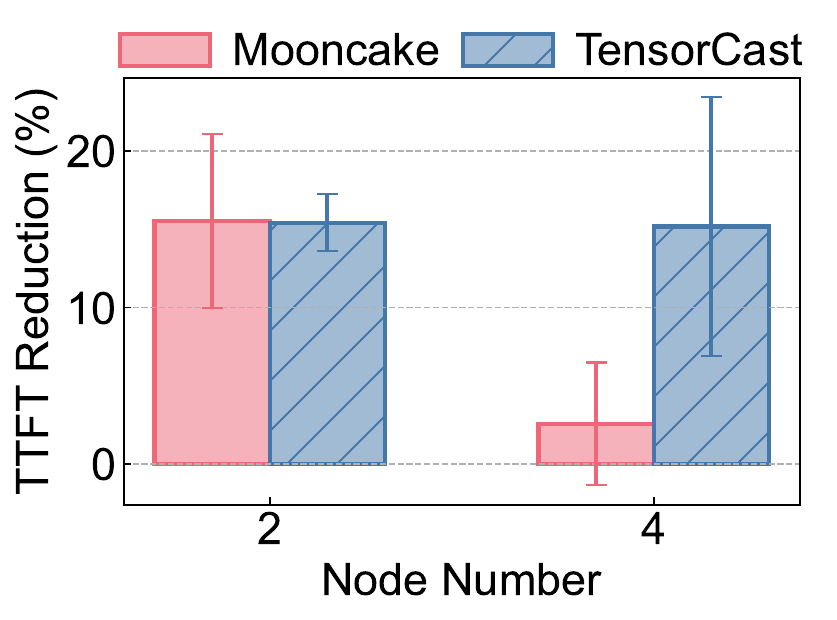}
		\caption{16k prompt w/ RDMA.}
		\label{fig:eval-kv-235b-16k-rdma}
	\end{subfigure}
	\hfill
	\begin{subfigure}[t]{0.24\linewidth}
        \centering
		\includegraphics[width=\linewidth]{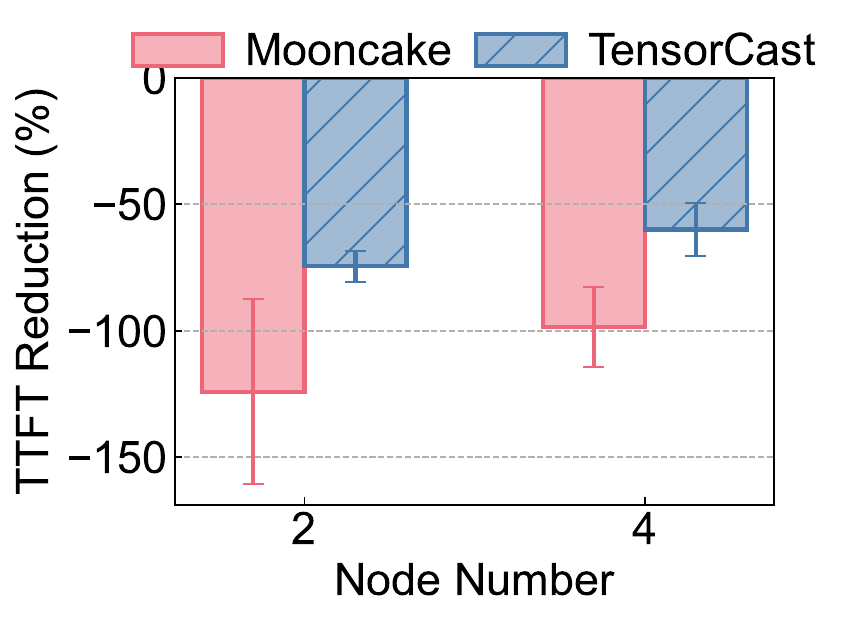}
		\caption{16k prompt w/o RDMA.}
		\label{fig:eval-kv-235b-16k-tcp}
	\end{subfigure}
    \hfill
	\begin{subfigure}[t]{0.24\linewidth}
        \centering
		\includegraphics[width=\linewidth]{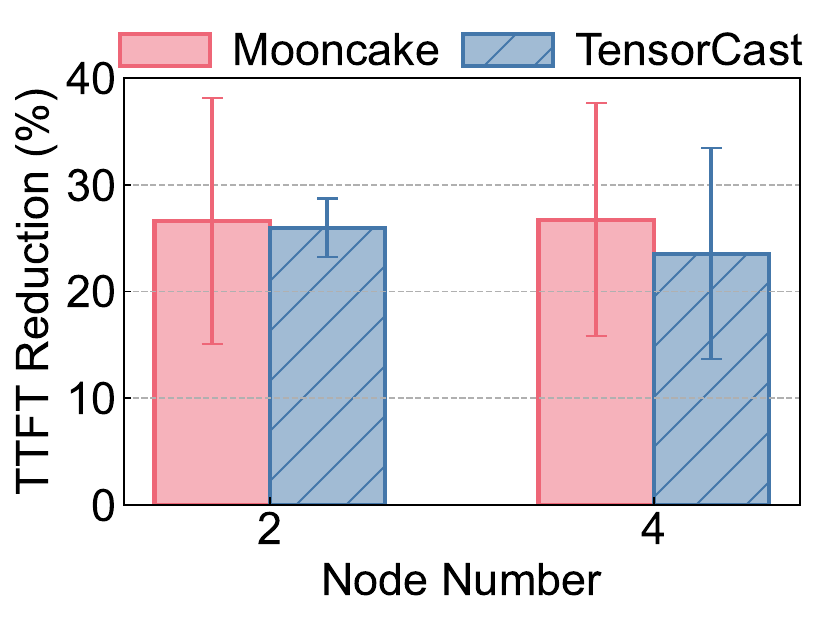}
		\caption{35k prompt w/ RDMA.}
		\label{fig:eval-kv-235b-35k-rdma}
	\end{subfigure}
	\hfill
	\begin{subfigure}[t]{0.24\linewidth}
        \centering
		\includegraphics[width=\linewidth]{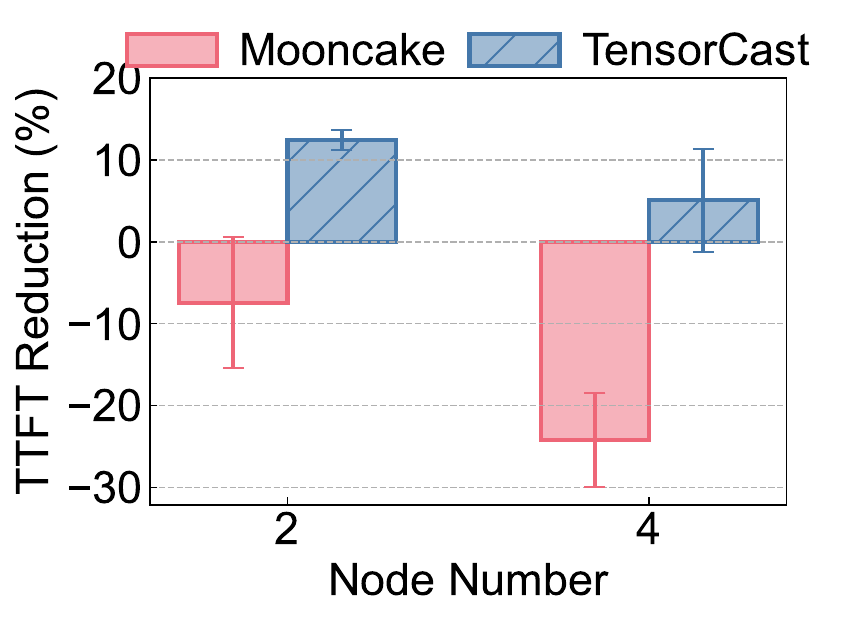}
		\caption{35k prompt w/o RDMA.}
		\label{fig:eval-kv-235b-35k-tcp}
	\end{subfigure}
    \caption{Qwen3-235B-A22B TTFT reduction ratio with \sys and Mooncake KV backend storage.}
    \label{fig:eval-kv-235b}
\end{figure*}

\subsection{Model Weight Materialization}\label{sec:eval-loader}
We evaluate \sys's ability to materialize large low-cardinality tensor
states during elastic LLM serving by integrating it into vLLM as a
model weight loading module.

\nosection{Setup.}
We use a single node with 64 CPUs, 500 GB
DRAM, and 8 $\times$ H800 GPUs to launch a vLLM instance (TP=8). This
setup emulates a MaaS auto-scaling scenario, where new instances are
launched as quickly as possible.
We load model weights in \texttt{safetensors} format from three storage
types:
\textit{i)} a high-performance distributed file system (JuiceFS
\cite{jfs});
\textit{ii)} node-local SSD; and
\textit{iii)} node-local DRAM (tmpfs).
Given a model path on each storage type, the vLLM instance materializes
weights through four configurations:
\textit{i) Default}, vLLM's native weight loader;
\textit{ii) InstantTensor} \cite{instenttensor}, a fast weight loader
with pipelined and zero-copy I/O;
\textit{iii) \sys cold}, where a \sys worker reads rank-specific weight
slices from storage and materializes them into VRAM, after which the
vLLM process consumes the tensors through the
\texttt{tensor\_dict} API;
and
\textit{iv) \sys warm}, where weight slices are pre-materialized locally
through the \texttt{prefetch} API and vLLM directly consumes the
materialized tensors.
This design exploits \sys's separation between tensor lifecycle
management and execution initialization, allowing tensor
materialization to occur concurrently with instance launch or during a
pre-flight stage.

\nosection{Results.}
We report both the model weight loading time during instance launch and the end-to-end instance-ready time to evaluate how \sys accelerates instance initialization. As shown in \figref{fig:eval-loader-30b} and \figref{fig:eval-loader-235b}, \sys consistently outperforms all baselines across settings. 
In particular, under JFS, a common case where MaaS providers host model weights, for the Qwen3-30B-A3B model, \sys-cold reduces model weight loading time by 60.7$\times$ and 10.2$\times$ compared to Default and InstantTensor, respectively (\figref{fig:eval-loader-30b-data_plane}). Consequently, it reduces end-to-end launch-ready time by 28.5$\times$ and 5.4$\times$, respectively (\figref{fig:eval-loader-30b-e2e}).
These gains stem from two factors: \textit{i)} \sys concurrently reads slice views for each rank while also issuing parallel read streams for each rank’s assigned slice; and \textit{ii)} tensor transfers to VRAM are pipelined to hide H2D copy overhead.
For \sys-warm, vLLM consumes pre-materialized tensors in less than 1 second, as \sys uses CUDA IPC to enable zero-copy access. As a result, end-to-end launch time is only dominated by runtime initialization overhead, yielding 228.6$\times$ and 40.7$\times$ speedups over Default and InstantTensor, respectively, for the Qwen3-235B-A22B model (\figref{fig:eval-loader-235b-e2e}).
These results demonstrate that \sys can efficiently materialize large
immutable tensor states while preserving a general tensor management
abstraction.

\subsection{Model Weight Synchronization}\label{sec:eval-updater}
We evaluate \sys's ability to synchronize versioned tensor states across distributed LLM instances by integrating it into SGLang as a new model weight updating module.

\nosection{Setup.} The experimental setup consists of
a two-node cluster, where each node runs a \sys worker with 64 CPUs,
500 GB DRAM, and 4 $\times$ H800 GPUs.
We use the \texttt{WeightPublisher} helper provided by \sys, which wraps
the \texttt{put()} API to publish different versions of model weights
into a \sys worker on one node. A serving SGLang instance on another
node synchronizes model weights by retrieving tensors from the
\sys cluster. This setup represents the training-serving synchronization
pattern commonly seen in RL post-training, where a trainer publishes
updated model weights and rollout workers periodically synchronize their
parameters.
During synchronization, each tensor-parallel (TP) rank retrieves only
its corresponding tensor slice through the \texttt{view()} API,
demonstrating \sys's ability to materialize different tensor layouts
from a shared tensor state. To ensure a fair comparison against the
\textit{Default} baseline, which retrieves full model weights from JFS
through a file path, we disable RDMA for \sys.

\nosection{Results.}
As shown in \figref{fig:eval-updater}, \sys accelerates model weight
synchronization by 1.14$\times$ to 2.63$\times$ compared with the
default updater. The acceleration trends vary with TP size and model
scale. For the Qwen3-14B model, increasing TP size introduces additional
tensor slicing overhead on the \sys worker, which becomes the dominant
cost compared with tensor transfer. In contrast, for the larger
Qwen3-32B model, \sys benefits from concurrently transferring multiple
tensor views, better utilizing available network bandwidth and achieving
higher acceleration as TP size increases.
These results demonstrate that \sys can efficiently manage versioned
tensor states across distributed execution instances while allowing
different consumers to materialize the tensor representations required
by their parallelism configurations.

\begin{table*}[t]
  \resizebox{\linewidth}{!}{%
  \begin{tabular}{lccccccp{8.5cm}}
  \toprule
  \textbf{Preset} & \textbf{$\mu$} & \textbf{$\sigma$} & \textbf{Median} & \textbf{Mean} & \textbf{P5} & \textbf{P95} & \textbf{Represents} \\
  \midrule
  fast & 2.1 & 0.6 & 8.2 s & 9.8 s & 3.0 s & 22.0 s & Tight coding-agent loop; short tool
  replies dominate \\
  medium & 3.0 & 0.8 & 20.1 s & 27.7 s & 5.4 s & 75.0 s & Typical SWE agent with mixed view, edit, and tool execution \\
  slow & 4.1 & 1.0 & 60.3 s & 99.5 s & 11.6 s & 311 s & Long-tool-dominated workflows; research/planning agents \\
  \bottomrule
  \end{tabular}
  }
  \ncaption{The three inter-turn delay presets used by the LogNorm sampling within a session.}
  \label{tab:lognorm-presets}
\end{table*}

\subsection{High-Cardinality Tensor Management}\label{sec:eval-kv}
We evaluate \sys's ability to manage high-cardinality tensor states by
integrating it into SGLang's \texttt{HiCache} module
\cite{sglang-hicache} as a KV cache storage backend. Each KV page is
represented as a high-cardinality \texttt{Artifact}, with KV pages distributed across \sys workers.

\nosection{Setup.}
The experimental setup consists of an $N$-node cluster, where each node
is equipped with 64 CPU, 1TB DRAM, 8 $\times$ H800 GPUs, and 200 Gbps
RDMA links. We deploy one \sys worker and one SGLang instance on each
node. To evaluate KV reuse, we first send prompts from the LongBench
dataset \cite{bai2024-longbench} to a single instance, and subsequently
resend the same prompts to all other instances in the cluster
simultaneously. This workload forces $N-1$ instances to concurrently
retrieve KV pages from the \sys cluster, guaranteeing a 100\% cache hit
rate and stress-testing high-cardinality tensor retrieval.
We group prompts into short and long categories, with approximately 16k
and 35k tokens on average, respectively. We compare \sys against the
state-of-the-art Mooncake \cite{qin2025-mooncake} integration in SGLang
under identical settings. Mooncake represents a highly optimized
KV-centric system, allowing us to evaluate whether \sys's general tensor
lifecycle abstraction introduces performance overhead compared with a
specialized KV management design.

\nosection{Results.}
We measure the average TTFT reduction of the subsequent $N-1$ instances
relative to the first instance to evaluate the effectiveness of KV reuse.
We also evaluate RDMA-disabled settings, since KV pages are increasingly
transferred across clusters or datacenters without RDMA interconnects
\cite{qin2026-prfaas}.
For the Qwen3-32B model with TP=2, as shown in
\figref{fig:eval-kv-32b}, with RDMA enabled
(\figref{fig:eval-kv-32b-16k-rdma} and
\figref{fig:eval-kv-32b-35k-rdma}), \sys achieves performance
comparable to Mooncake, with TTFT reductions ranging from 60\% to 87.5\%
(\ie 2.5$\times$ to 8$\times$ speedup). Performance improves as the
number of nodes $N$ increases because \sys distributes high-cardinality
tensor states across shard homes rather than centralizing metadata
management through the global store (\secref{sec:arch-tensor-pool}).
With RDMA disabled (\figref{fig:eval-kv-32b-16k-tcp} and
\figref{fig:eval-kv-32b-35k-tcp}), \sys significantly outperforms
Mooncake while maintaining positive TTFT reductions. This improvement
comes from \sys's userspace mTCP implementation
\cite{jeong2014-mtcp} with multipath transmission, which reduces kernel
copy overhead and aggregates higher effective bandwidth compared with
in-kernel TCP.
For the Qwen3-235B model with TP=8, as shown in
\figref{fig:eval-kv-235b}, \sys again achieves comparable performance
to Mooncake under RDMA and better performance without RDMA. Compared
with TP=2, TP=8 introduces 4$\times$ more concurrent KV page retrieval
requests, increasing contention and reducing TTFT improvements for both
systems.
These results demonstrate that \sys can efficiently manage highly
dynamic, high-cardinality tensor states while preserving the performance
of specialized KV cache systems.

\begin{figure*}[t]
    \centering
    \includegraphics[width=\linewidth]{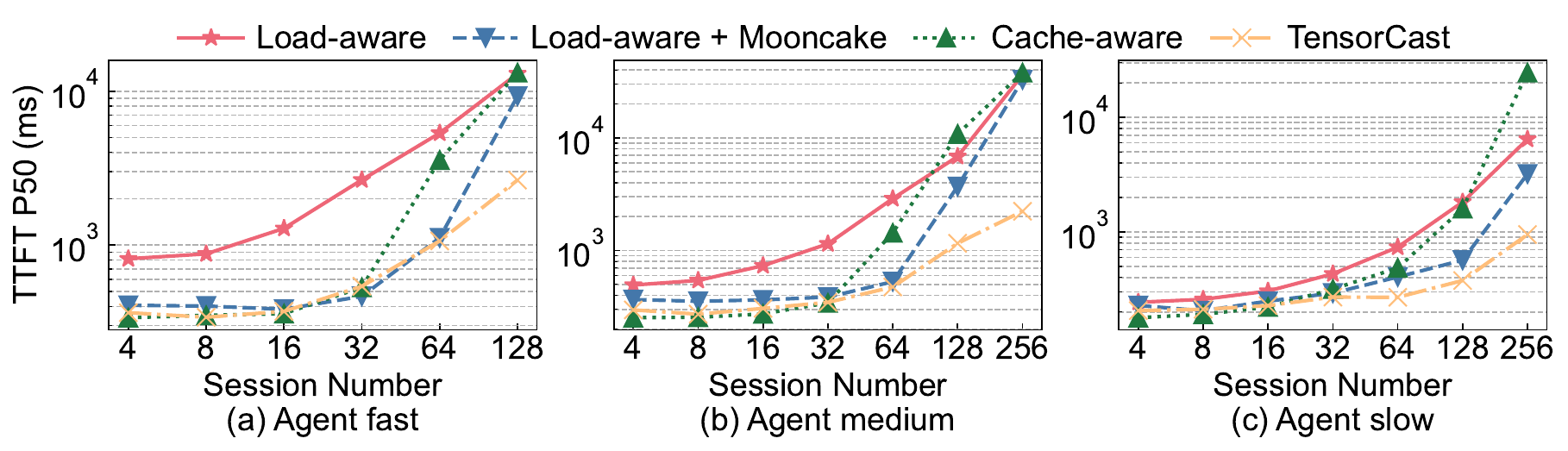}
    \caption{Request TTFT against different router policies with three levels of agentic coding workloads.}
    \label{fig:eval-router-ttft}
\end{figure*}

\begin{figure*}[t]
    \centering
    \includegraphics[width=\linewidth]{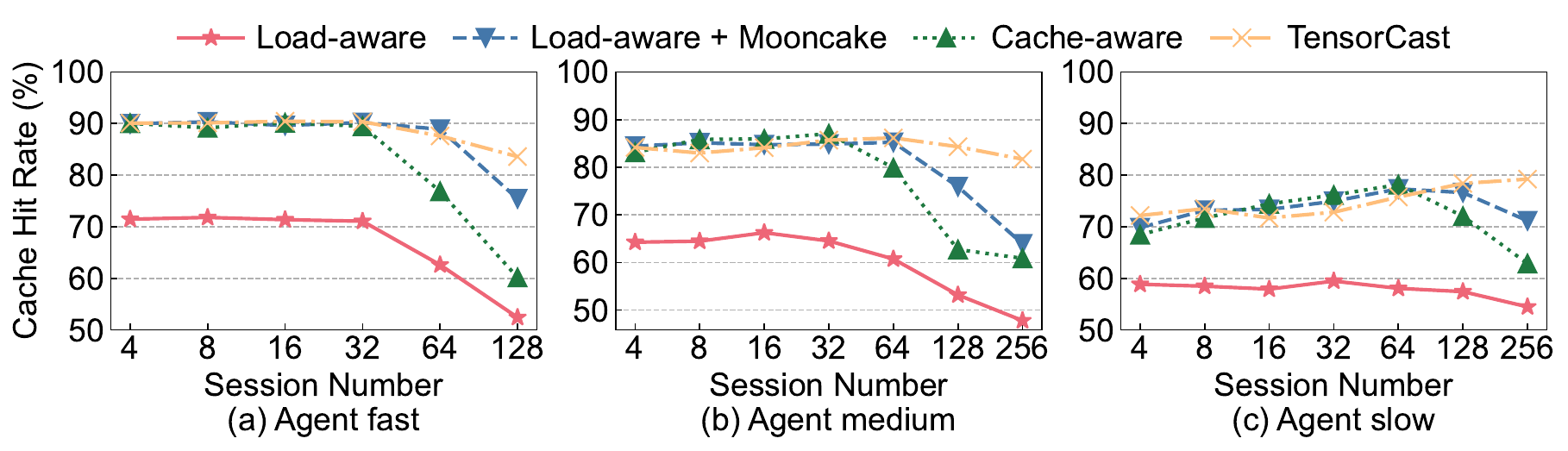}
    \caption{Request cache hit rate against different router policies with three levels of agentic coding workloads.}
    \label{fig:eval-router-cache}
\end{figure*}

\subsection{Programmable Request Router}\label{sec:eval-router}
We evaluate \sys's programmability by implementing a workload-specific
request routing policy that composes multiple tensor lifecycle
operations. We integrate an instance adaptor into SGLang, enabling callers to execute the instance operations listed in
\tabref{tab:apis} on SGLang instances.

\nosection{Setup.}
Based on the widely used cache-aware routing policy, we implement a lightweight rebalancing policy that periodically observes instance states and migrates selected requests, together with their KV caches, from the most-loaded instance to the least-loaded instance using the \texttt{Plan} abstraction shown in
\lstref{code:tc-router} (see detailed rebalancing policy in
\secref{sec:appendix-rebalance-policy}). This combined policy aims to
preserve KV cache locality while avoiding load imbalance across
instances, without losing context during migration. The rebalancing
policy is independent of \sys and can be customized without modifying
the SGLang scheduler, KV cache backend, or \sys runtime.
We compare the \sys router against three baseline routers implemented in
the SGLang model gateway \cite{sgl-model-gateway}:
\textit{i)} a pure load-aware router using the widely used
power-of-two-choices policy \cite{mitzenmacher2002-p2c};
\textit{ii)} a load-aware router with Mooncake as the backend for
sharing KV caches across instances; and
\textit{iii)} a pure cache-aware router that routes requests to
instances with the longest local prefix match. The cluster consists of
four nodes, each equipped with 2 $\times$ H20 GPUs. Each node runs one
Qwen3-32B instance with TP=2.

\nosection{Workloads.}
We generate dynamic multi-turn workloads from real agent rollouts on the
SWE-Gym task pool, collected by running the OpenHands agent on
approximately 2,400 real-world Python repository tasks
\cite{swe-gym-openhands}. We treat each rollout trajectory as a
multi-turn request session and maintain concurrent sessions as workload
traffic to the cluster. The interval between consecutive requests within
a session is sampled from a LogNormal distribution, since this
right-skewed distribution reflects human-action intervals
\cite{brown2005-statistical, malmgren2008-poissonian}. We use three
LogNormal parameter presets--fast, medium, and slow--with parameters
$(\mu, \sigma)$ as described in \tabref{tab:lognorm-presets}.

\nosection{Results.}
We measure the median TTFT of all requests across all instances and
report the results in \figref{fig:eval-router-ttft}. As the number of
concurrent user sessions increases, the median TTFT increases rapidly
for all methods. However, the rebalancing policy implemented by
composing \sys lifecycle APIs achieves the lowest median TTFT across all
workloads. Specifically, under the heaviest loads, 128 sessions for
\textit{fast} and 256 sessions for \textit{medium} and \textit{slow},
the \sys router reduces median TTFT by 71.4\%, 93.2\%, and 70.4\% for
\textit{fast}, \textit{medium}, and \textit{slow}, respectively,
compared with the load-aware router using Mooncake.
We also report the average cache hit rate in
\figref{fig:eval-router-cache}. As the number of concurrent user
sessions increases, the cache hit rate decreases for all methods due to
limited local cache capacity on each instance and Mooncake overload. In
contrast, the \sys router exhibits the smallest hit-rate drop by jointly
balancing KV cache placement and instance load while preserving
requests from the same session on cache-hitting instances. This
behavior effectively reduces request TTFT.
These results validate the programmability of \sys: developers can
compose tensor lifecycle primitives to implement new optimization
policies that coordinate request scheduling and tensor placement,
without modifying existing serving components.

\section{Discussion}\label{sec:discussion}

\nosection{Applicability beyond inference workloads.}
While our integration and evaluations in \secref{sec:eval} focus on inference workloads, \sys can also express tensor lifecycles in training workloads: At the abstraction level, checkpoints can be mapped to versioned, system-owned \texttt{Artifact}s, with \texttt{view()} and \texttt{transform\_into()} expressing rank-specific resharding; gradients, optimizer states, and long-lived activations can be exposed as caller-leased tensors. We leave the integration of \sys with training frameworks like Megatron-LM \cite{shoeybi2019-megatron} and DeepSpeed \cite{rasley2020-deepspeed} for future work.

\nosection{Integration cost and abstraction overhead.}
Decoupling tensor management does not remove engine-specific integration: \sys confines the export, import, or transformation of engine-resident tensor states to an instance adaptor, so customized programs can evolve in callers without repeatedly modifying engines, transports, or storage backends. We do not claim that a general abstraction is inherently faster than every specialized implementation. \secref{sec:eval-loader}--\secref{sec:eval-kv} evaluate reusable mechanisms and show that \sys achieves comparable or better performance, while \secref{sec:eval-router} shows optimization enabled by composition.

\section{Other Related Work}
In addition to the studies for efficient tensor management discussed in \secref{sec:bkgd}, many other studies have investigated optimizations for LLM workloads.

\nosection{Efficient LLM inference.}
An increasing body of literature focuses on enhancing the efficiency of Transformer-based LLM inference. These optimizations encompass parallelism \cite{zheng2022alpa, aminabadi2022deepspeed, wu2024loongserve}, request scheduling \cite{wu2023fastserve, yu2022orca, agrawal2024sarathi}, memory management \cite{kwon2023-pagedAttention, lee2024infinigen, zhang2025jenga}, quantization \cite{wang2023bitnet, lin2024awq}, and specialized GPU kernels \cite{dao2022flashattention, ye2025flashinfer}. Serving as a complementary approach, \sys offers flexible, large-scale tensor management and integrates seamlessly with these existing techniques.

\nosection{Programmable LLM systems.}
Numerous studies explore the programming of LLM behaviors. One line of research targets frameworks for prompt engineering and agent orchestration, including LangChain \cite{langchain}, LangFlow \cite{langflow}, AutoGen \cite{wu2023-autogen}, and DSPy \cite{khattab2024-dspy}. These tools simplify agentic workflows to accelerate LLM application development. Another direction provides fine-grained control over the LLM inference process--encompassing tokenization, attention, KV cache management, and decoding--through systems like microserving LLM \cite{jin2024-llm-microserving} and PIE \cite{gim2025-pie}. However, existing programmable LLM systems operate primarily at the \textit{execution layer} of LLM workloads. In contrast, \sys introduces programmable management of the data lifecycle (\ie tensors), making it an orthogonal and complementary approach to current frameworks.

\section{Conclusion Remarks}

As LLM systems continue to evolve toward increasingly dynamic and
stateful workloads, efficient tensor lifecycle management becomes a
fundamental infrastructure challenge beyond individual optimization
techniques. \sys introduces the missing tensor management layer
by decoupling tensor states from the computation that produces and
consumes them, enabling tensors to be managed as programmable,
first-class system resources. Through lifecycle primitives and a
distributed runtime, \sys achieves the efficiency of specialized
tensor management mechanisms while allowing developers to compose new
workload-specific policies. Our results demonstrate that future LLM
infrastructure should not only optimize computation over tensors, but
also provide programmable control over the lifecycle of the tensor
states that drive these computations.

    \newpage

    \balance{
        \bibliographystyle{unsrt}
        \bibliography{ref}
    }

    \newpage

    \appendix

\section{Router Rebalancing Policy}\label{sec:appendix-rebalance-policy}

\begin{algorithm}[t]
      \caption{Router Request Rebalancing Policy.}
      \label{alg:rebalance-policy}

      \textbf{For Each Rebalance Tick:}

      \ForEach{instance $i$} {
          $L_i \leftarrow \alpha \cdot \text{Load}(i) + (1-\alpha)L_i$
          \Comment{Update instance load using EWMA}
      }

      $src \leftarrow \arg\max_i L_i$

      $dst \leftarrow \arg\min_{i \neq src} L_i$

      $gap \leftarrow L_{src} - L_{dst}$

      $ratio \leftarrow L_{src} / \max(L_{dst}, 1)$

      \If{$gap < \theta_{abs}$ \textbf{or} $ratio < \theta_{rel}$} {
          \textbf{return} $\emptyset$ \Comment{No rebalance}
      }

      $R \leftarrow \text{EligibleRequests}(src)$ \Comment{Filter out running or migrating requests}

      \If{$R = \emptyset$} {
          \textbf{return} $\emptyset$
      }

      \ForEach{req $r \in R$} {
          $active_r \leftarrow e^{-(now - last\_active_r)/H}$

          $penalty_r \leftarrow 1 / (1 + \lambda \cdot migrations_r)$

          $score_r \leftarrow token_r \cdot active_r \cdot penalty_r$
      }

      $r^* \leftarrow \arg\max_{r \in R} score_r$

      \textbf{return} $(r^*, src, dst)$ \Comment{Migrate one session}
\end{algorithm}

As shown in \algoref{alg:rebalance-policy}, the programmable router in \secref{sec:eval-router} runs a simple rebalancing policy periodically. At each tick, it first updates an exponentially weighted moving average (EWMA) load estimate for each instance using the current queue depth, including both waiting and running requests. It then selects the most-loaded instance as the source and the least-loaded instance as the target. A rebalance is triggered only when both the absolute load gap and the relative load ratio exceed predefined thresholds, $\theta_{abs}$ and $\theta_{rel}$, thereby avoiding unnecessary migrations when the cluster is already balanced.

When rebalancing is needed, the policy considers only eligible requests on the source instance, excluding requests that are currently running or already under migration. Each eligible request is scored using three factors: request length in tokens, which estimates the potential KV reuse benefit; recency, which favors recently active requests; and a history penalty, which discourages repeatedly migrating the same request and avoids ping-pong behavior. The request with the highest score is selected as the migration candidate from the source to the target.

\end{document}